\documentclass[aps,prd,amsmath,amssymb,amsfonts,floats,floatfix,superscriptaddress,nofootinbib,notitlepage]{revtex4-1}
\usepackage{graphicx}
\usepackage{graphics}
\usepackage[retainorgcmds]{IEEEtrantools}
\usepackage{color}
\usepackage{multirow}
\usepackage{xspace} 
\usepackage{slashed}
\usepackage[colorlinks=true, citecolor=blue,urlcolor=blue]{hyperref}
\usepackage{graphicx,marvosym,subfigure,amssymb,amsmath}
\usepackage{comment}
\usepackage{graphicx}

\DeclareMathAlphabet{\mathpzc}{OT1}{pzc}{m}{it}
\usepackage{calligra}
\usepackage[cal=boondox, calscaled=.94, bb=ams, frak=mma, frakscaled=.97, scr=rsfs]{mathalfa} %

\usepackage[top=1in, bottom=1in, left=.9in, right=.9in]{geometry}
\newcommand{\comm}[1]{}
\newcommand{\PZ}[1]{} 
\newcommand{\MC}[1]{}

\newcommand{\indmode}{\ell m\omega} 
\newcommand{\indmodeocc}{\ell m\omega^*}
\newcommand{\indmodemmmocc}{\ell, -m,-\omega^*}
\newcommand{\indmodemmmo}{\ell, -m,-\omega}

\newcommand{\q}[2]{q_{#1}^{#2}}
\newcommand{\e}{\chi_{s}}
\newcommand{\et}{ \chi_{-s}}
\newcommand{\eet}{ \chi_{\pm s}}
\newcommand{\chics}{\chi_{s\vert c}}
\newcommand{\chicms}{\chi_{-s\vert c}}

\newcommand{\aaa}{d}
\newcommand{\aak}{\bar \aaa}

\newcommand{\ocr}{m}
\newcommand{\eps}{\epsilon}
\newcommand{\epsc}{\epsilon_c}

\newcommand{\Rin}{R^{\text{in}}_{\indmode}}
\newcommand{\Rup}{R^{\text{up}}_{\indmode}}
\newcommand{\Rout}{R^{\text{out}}_{\indmode}}
\newcommand{\Rinhat}{\hat{R}^{\text{in}}_{\indmode}}
\newcommand{\Ruphat}{\hat{R}^{\text{up}}_{\indmode}}
\newcommand{\Rinhatccmmmocc}{\hat{R}^{\text{in}^*}_{\indmodemmmocc}}
\newcommand{\Ruphatccmmmocc}{\hat{R}^{\text{up}^*}_{\indmodemmmocc}}
\newcommand{\Ruphatcc}{\hat{R}^{\text{up}^*}_{\indmode}}
\newcommand{\Routhat}{\hat{R}^{\text{out}}_{\indmode}}

\newcommand{\Rpnu}{R^{\nu}_{+}}
\newcommand{\Rmnu}{R^{\nu}_{-}}
\newcommand{\Rpnuhat}{\hat{R}^{\nu}_{+}}

\newcommand{\Cinf}{\infty}

\newcommand{\Rininc}{\mathcal{I}_{\text{in}}}
\newcommand{\Rinref}{\mathcal{R}_{\text{in}}}
\newcommand{\Rintra}{\mathcal{T}_{\text{in}}}
\newcommand{\Routtra}{\mathcal{T}_{\text{out}}}
\newcommand{\Rupinc}{\mathcal{I}_{\text{up}}}
\newcommand{\Rupref}{\mathcal{R}_{\text{up}}}
\newcommand{\Ruptra}{\mathcal{T}_{\text{up}}}
\newcommand{\Rininchat}{\mathcal{\hat I}_{\text{in}}}
\newcommand{\Rinrefhat}{\mathcal{\hat R}_{\text{in}}}
\newcommand{\Rinrefhatcc}{\mathcal{\hat R^*}_{\text{in}}}
\newcommand{\Rupinchat}{\mathcal{\hat I}_{\text{up}}}
\newcommand{\Ruprefhat}{\mathcal{\hat R}_{\text{up}}}
\newcommand{\Ruprefhatcc}{\mathcal{\hat R^*}_{\text{up}}}
\newcommand{\Rpnutra}{\mathcal{T}_{+}}
\newcommand{\Rmnutra}{\mathcal{T}_{-}}

\newcommand{\W}{\mathcal{W}}
\newcommand{\Wbp}{\hat \W^+}
\newcommand{\Wbm}{\hat \W^-}
\newcommand{\What}{\hat \W}
\newcommand{\Whatpm}{\hat \W^{+/-}}
\newcommand{\Wcritp}{\hat{\bar \W}^+}
\newcommand{\Wcritm}{\hat{\bar \W}^-}
\newcommand{\Wcritpm}{\hat{\bar \W}^{+/-}}

\newcommand{\ob}{k }
\newcommand{\fNIA}{\sigma}
\newcommand{\fcrit}{\bar \sigma}

\newcommand{\aMST}[1]{a^{\nu}_{#1}}
\newcommand{\aMSTmnu}[1]{a^{-\nu-1}_{#1}}
\newcommand{\bMST}[1]{b^{\nu}_{#1}}
\newcommand{\bMSTmnu}[1]{b^{-\nu-1}_{#1}}

\newcommand{\SKnu}{\mathcal{S}_{\nu_c}}
\newcommand{\SKmnu}{\mathcal{S}_{-\nu_c-1}}

\newcommand{\Disc}{\delta}
\newcommand{\Discrit}{\bar \delta}
\newcommand{\glmo}{\tilde{g}_{\indmode}}
\newcommand{\glmoccmmmocc}{\tilde{g}^*_{\indmodemmmocc}}
\newcommand{\Glm}{\mathsf G_{\ell m}}

\newcommand{\zetainfp}{\zeta^{(\infty)}_{+}}
\newcommand{\zetahorp}{\zeta^{(0)}_{+}}
\newcommand{\zetahorm}{\zeta^{(0)}_{-}}

\newcommand{\rArbit}{p}

\newcommand{\eigenl}{{}_s\lambda_{\indmode}}

\newcommand{\eigenA}{{}_sA_{\indmode}}
\newcommand{\eigenAms}{{}_{-s}A_{\indmode}}
\newcommand{\eigenAmmmo}{{}_{s}A_{\ell,- m,-\omega}}
\newcommand{\eigenAcco}{{}_{s}A_{\ell  m\omega^*}}
\newcommand{\eigenbA}{{}_s\bar{A}_{\indmode}}
\newcommand{\eigenAmeqo}{{}_sA_{\ell,m,m}}
\newcommand{\eigenK}{{}_sK_{\ell m}}
\newcommand{\eigenKms}{{}_{-s}K_{\ell m}}

\newcommand{\eigenSS}{\mathcal{Z}}

\newcommand{\nn}{\nonumber}

\begin{document}
\global\parskip 6pt

\author{Marc Casals}
\email{mcasals@cbpf.br, marc.casals@ucd.ie.}
\affiliation{Centro Brasileiro de Pesquisas F\'isicas (CBPF),  Rio de Janeiro, 
CEP 22290-180, 
Brazil.}
\affiliation{School of Mathematics and Statistics,
University College Dublin, Belfield, Dublin 4, Ireland}

\author{Peter Zimmerman}
\email{peter.zimmerman@aei.mpg.de}
\affiliation{Max Planck Institute for Gravitational Physics (Albert Einstein Institute)\\
  Am M\"uhlenberg 1, 14476 Potsdam, Germany}

\title{Perturbations of Extremal Kerr Spacetime: \\ Analytic Framework and Late-time Tails}

\begin{abstract} 
We develop a complete and systematic analytical approach to field perturbations of  extremal Kerr spacetime based on the 
formalism 
of Mano, Suzuki and Takasugi (MST)
for the Teukolsky equation. Analytical expressions for the radial solutions and frequency-domain Green function in terms of infinite series of special functions are presented.  
As an application, we compute, for the first time, the leading late-time behavior due to the branch point at zero frequency of scalar, gravitational, and electromagnetic field perturbations on and off the event horizon. We also use the MST method to compute the leading behavior of the Green function modes near the branch point at the superradiant bound frequency and show that this behavior agrees with existing results in the literature using a different method.
\end{abstract}

\date{\today}
\maketitle

\tableofcontents

\section{Introduction}
\MC{The table of contents won't be allowed by PRD}
The prominence of black holes in modern physics and astrophysics makes study of their perturbations of essential importance.   Regge, Wheeler, Zerilli, and Moncrief \cite{Regge:1957td,PhysRevLett.24.737,Moncrief:1974am} 
pioneered work
on linear field perturbations of spherically symmetric (Schwarzschild) black holes.
The astrophysically-relevant case, however, is that
of rotating black holes, which are described by
the Kerr metric.
The rotating case was cracked by Teukolsky \cite{Teukolsky:1973ha}, who derived a  master equation for scalar (spin-0), fermion (spin-1/2), electromagnetic (spin-1), and gravitational (spin-2) perturbations of the Kerr metric.  While Teukolsky's equation can be solved numerically (say, in the time domain), it is of interest to develop complementary analytical techniques.  
Teukolsky's master equation is a partial differential equation which separates into 
a radial and an angular ordinary differential equation by going into the frequency domain.
By adopting 
such a frequency domain approach,
Leaver~\cite{Leaver:1986a} found analytical solutions of the radial equation in terms of infinite series involving special functions. Mano, Suzuki, and Takasugi (MST)~\cite{Mano:Suzuki:Takasugi:1996,Mano:1996mf} cleverly reformulated Leaver's solutions to produce a practical method of computing observable quantities such as the gravitational waveform from 
the inspiral of a compact object into a supermassive black hole in the
extreme mass-ratio regime.  With the advent of computer algebra programs capable of efficiently manipulating and computing special functions, this ``MST method'' has gained in popularity to become competitive with, and in many ways superior to, direct numerical solution of the linearized perturbation equation.
The MST method has been used for calculating the
self-force, post-Newtonian coefficients and gauge-invariant
quantities, the retarded Green function, the quantum correlator, the renormalized expectation value of the quantum stress-energy tensor and radiation emission in  Schwarzschild and Kerr spacetimes in~\cite{van2016gravitational,zhang2013quasinormal,CDOW13,BUSS2018168,levi2017renormalized,bini2014high}; of particular relevance to this paper, it has also been used to calculate the late-time tail to high-order  in Schwarzschild and Kerr spacetimes in~\cite{Casals:Ottewill:2015,casals2016high}.

Kerr black holes possess an outer event horizon and an inner Cauchy horizon beyond which the Cauchy value problem is not well-posed.
In the case of maximal rotation, called extremal, the Cauchy and event horizons coincide.
Since the original formulation for spin-field perturbations of vacuum, asymptotically flat, non-extremal black holes~\cite{Mano:Suzuki:Takasugi:1996,Mano:1996mf} (compiled in a review in~\cite{Sasaki:2003xr}), the MST method has been extended in a variety of ways to encompass electrically-charged black holes and/or a nonzero cosmological constant~\cite{suzuki1999analytic,suzuki1998perturbations,suzuki2000absorption}.  However, to the best of our knowledge, there has been no prior MST work on extremal black holes.  The extension is not straightforward since the coincidence of the inner and outer horizons converts a pair of regular singular points of the radial equation into a single irregular singular point, changing the character of the series solutions.  We side-step the difficulty by starting with a functional expansion adapted to extremal Kerr and develop a Leaver-MST method accordingly.  The MST-type series that we derive in extremal Kerr provides a practical and efficient formulation for analytic evaluation of integer-spin perturbations of this spacetime.

As an application, we compute the late-time behavior (``tail'') of the perturbing field.  In the nonextremal case, the tail arises from a branch point that the radial
solutions possess at the origin of the complex frequency plane (i.e., at zero frequency)~\cite{Leaver:1986,PhysRevLett.84.10,casals2016high}.  In the extremal case there is an additional branch point at the so-called superradiant bound frequency that must also be considered~\cite{PhysRevD.64.104021}. In previous work \cite{Casals:2016mel, Gralla:2017lto}, we have computed the extremal Kerr tail from the branch point at the superradiant bound frequency, showing that the asymptotic decay of the perturbing field, whether it be metric, vector potential, or scalar, is $1/v$ off the horizon and $1/\sqrt{v}$ on the horizon ($v$ being advanced time). The difference in the rates on and off the horizon accounts for the divergent growth of transverse derivatives at the horizon \cite{aretakis2012horizon, lucietti2012gravitational}, a phenomenon named after its discoverer, Aretakis \cite{Aretakis:2011ha, Aretakis:2011hc}.  In these calculations we used the method of matched asymptotic expansions (MAE) to compute the leading late-time behavior, which comes from the behavior of the modes near the superradiant bound frequency. In this work we show that the transfer function (i.e., the fixed frequency modes of the retarded Green function) used in MAE calculations is recovered exactly by the leading-order term in the MST series that we derive.  This unites previously distinct techniques, provides a more rigorous justification for the MAE, and shows how it can be systematically corrected to arbitrary order in frequency. Moreover, in the aforementioned MAE calculations the late-time rates due to the superradiant bound frequency were reported under the assumption that the tail due to the branch point at the origin is subleading.  Here we justify this assumption, showing that the tail from the origin is in indeed subleading at the horizon, going as
$v^{-3-2\ell}$
along the future event horizon (see Eq.~\eqref{eq:delta Glm,v inf,x->0}). We also derive asymptotic decay rates at future null and timelike infinity. We find that these rates are:
$u^{-2+s-\ell}$ along 
future 
null
infinity
(see Eq.~\eqref{eq:delta Glm,t inf}) 
and $t^{-3-2\ell}$
at future timelike infinity
(see Eq.~\eqref{eq:delta Glm,t inf x finite}), where $t$ is Boyer-Lindquist time, $u$ is retarded time
and $\ell$ is the multipole number in the decomposition in angular functions (i.e., spin-weighted spheroidal harmonics~\cite{Berti:2005gp,berti2006erratum}).
Our results are for nonaxisymmetric, integer-spin field perturbations\footnote{The axisymmetric case is already considered in~\cite{Casals:2016mel, Gralla:2017lto} with the exception of the tail at future null infinity. Also, it should be straightforward to extend our results to half-integer  values for the spin of the field.}.

The rest of this paper is organized as follows. In Sec.~\ref{sec:overview} we introduce the Teukolsky equation and its retarded Green function.
In Sec.~\ref{sec:MST} we develop the MST formalism for extremal Kerr.
In Sec.~\ref{sec:BC origin} we apply the MST formalism to obtain the formal contribution to the Green function from the branch cut down from the origin and derive the corresponding leading-order late-time tail.
In Sec.~\ref{sec:BC k=0} we  obtain the formal contribution to the Green function from the branch cut down from the superradiant bound frequency.
In Sec.~\ref{sec:link to MAE} we show that, in the limit to the superradiant bound frequency, the MST method recovers the MAE results.
In App.\ref{sec:App radial Leaver} we give Leaver's~\cite{Leaver:1986a} original expressions and relate them to ours.
 
We follow the notation and units of Sec.VIII.B of Ref.~\cite{Leaver:1986a}. In particular, we choose $c=G=1$ and the unusual choice $M=1/2$ for the mass of the black hole. 

\section{Perturbations of extremal Kerr}\label{sec:overview}

\subsection{Retarded Green function}

Scalar (spin $s=0$), electromagnetic ($s=\pm 1$), and gravitational ($s=\pm $2) perturbations
\footnote{Fermion ($s=\pm 1/2$) field perturbations also obey the Teukolsky equation. However, in this paper we assume integer $s$, which simplifies some of the formulas.} 
$\Psi$ of an extremal Kerr black hole are governed by a single ``master'' equation first derived by Teukolsky~\cite{Teukolsky:1973ha}. 
This is a ($3+1$)-dimensional, second-order wave equation.
Our main study concerns the
retarded Green function 
of this equation, 
where $x^\mu$ 
and $x^\mu{}'$ are spacetime points\footnote{In a common abuse of notation, we use the same symbol to denote spacetime points and their coordinates.}.
The retarded Green function $G(x^\mu,x^\mu{}')$  is
defined to vanish when $x^\mu$ is outside the causal future of  $x^\mu{}'$. 
We employ Boyer-Lindquist coordinates
$\{t\in\mathbb{R},r\in (M,\infty),\theta\in [0,\pi],\phi\in [0,2\pi)\}$ 
outside the event horizon of the black hole
and install the Kinnersley tetrad~\cite{Kinnersley1969}.
We denote the mass of the black hole by $M$ and let $r_H:=M$ denote the radius of the
event horizon at extremality.
Henceforth we choose units such that $M=1/2$, and so $r_H=1/2$.
Instead of the Boyer-Lindquist radial coordinate $r$ we shall use the shifted radial coordinate
\begin{align}
x := r-r_H = r-1/2 \in (0,\infty). 
\end{align}
In the shifted Boyer-Lindquist coordinates, the retarded Green function satisfies the fundamental equation~\cite{Teukolsky:1973ha}
\begin{equation}
    \mathcal O[G(x^\mu,x^\mu{}')] = \delta(t-t')\delta(x-x')\delta(\cos\theta-\cos\theta')\delta(\phi-\phi'),
\end{equation}
where $\mathcal O$ is the Teukolsky operator.
In the  metric signature $(-+++)$ that we use, $\mathcal O$ corresponds to minus the operator in the left-hand side of Eq.~(4.7) of~\cite{Teukolsky:1973ha}.

In order to calculate the retarded Green function,
we mode decompose into spin-weighted spheroidal harmonics ${}_sS_{\indmode}$~\cite{Berti:2005gp,berti2006erratum} and make use of the axisymmetry and stationarity of the spacetime.
Explicitly, we decompose $G$ as
\begin{align}
G(x^\mu,x^\mu{}')=- 
\frac{x'{}^{2s}}{2\pi}  \sum_{\ell=|s|}^{\infty}\sum_{m=-\ell}^{\ell}\int_{-\infty+i c}^{\infty+i c}\, 
e^{-i \omega t+i m\phi}
{}_s\eigenSS_{\indmode}(\theta,\theta')
\tilde g_{\indmode}(x,x') d\omega \,, \label{Eq:TDGF}
\end{align}
where 
\begin{equation}\label{eq:eigenSS}
{}_s\eigenSS_{\indmode}(\theta,\theta'):= {}_sS_{\indmode}(\theta){}_sS_{\indmode}^{*}(\theta'),    
\end{equation}
and $c>0$ ensures the integration contour of the inverse Laplace transform is in the analytic region of the transfer function
$\tilde g_{\ell m \omega}$. 
By the symmetries of the Kerr spacetime, we have set $t'=0$ and $\phi'=0$ without
loss of generality.

The spin-weighted spheroidal harmonics ${}_sS_{\indmode}$ are  understood to be evaluated at extremality, i.e., for  black hole angular momentum  $a$ per unit mass equal to the mass, i.e., $a=M=1/2$.
These angular functions satisfy the following ordinary differential equation:
\begin{equation} \label{eq:ang. teuk. Eq}
\left(
\frac{1}{\sin\theta}\frac{d}{d\theta}\left(
\sin\theta
\frac{d }
{d\theta}
\right)+
\frac{\omega^{2}\cos^2\theta}{4}
-\frac{m^2}{\sin^2\theta}
-
\omega s\cos\theta
-\frac{2ms\cos\theta}{\sin^2\theta}
-s^2\cot^2\theta+s+\eigenA
\right)
{}_sS_{\indmode}(\theta)=0,
\end{equation}
where $\eigenA$ is a separation constant.
Together with the boundary conditions of regularity at $\theta=0$ and $\pi$, this equation poses
an eigenvalue problem with
eigenvalue $\eigenA$.
For real frequencies, the spin-weighted spheroidal harmonics form a strongly complete set of eigenfunctions, whereas for complex frequencies they only form a weakly complete set~\cite{stewart1975stability}.
Following~\cite{Teukolsky:1973ha}, it is also convenient to define the quantity $\eigenl:=\eigenA + M^2 \omega^2 - 2 M m \omega=\eigenA + \omega^2/4 - m \omega$.
Our convention is to normalize the spin-weighted spheroidal harmonics such that 
\begin{equation}\label{eq:S normalization}
 \int_0^{2 \pi}  \! \! \int_0^\pi e^{i(m-m')\phi} {}_sS_{\indmode}(\theta) \,{}_s S_{\ell' m'\omega}(\theta) \sin\theta \, d\theta\, d\phi = 2 \pi \delta_{\ell \ell'}\delta_{m m'}.
\end{equation}

We now give
some useful properties of the angular eigenfunctions and their eigenvalues.
From the angular equation,
\eqref{eq:ang. teuk. Eq}
the following symmetries are manifest:
\begin{equation}\label{eq:symms eigen}
\eigenA+s=\eigenAms-s,\quad
\eigenA^*=\eigenAcco,\quad
\eigenA=\eigenAmmmo,
\end{equation}
\MC{I wonder whether the 2nd property above is also true when $\omega$ lies on an angular BC}
for the angular eigenvalue and
\begin{equation}\label{eq:symms eigenfunc}
{}_s\eigenSS_{\indmode}(\theta,\theta')=
{}_{-s}\eigenSS_{\indmodemmmo}(\theta,\theta')=
{}_{s}\eigenSS_{\indmodemmmo}(\pi-\theta,\pi-\theta'),\quad
{}_s\eigenSS^*_{\indmode}(\theta,\theta')=
{}_{s}\eigenSS_{\indmodeocc}(\theta,\theta'),
\end{equation}
for the angular eigenfunction product.

In its turn,
routine separation of variables reveals 
that the transfer function
obeys the
ordinary differential equation
\begin{equation} \label{eq:radial teuk. Eq}
\mathcal L[ \tilde g_{\ell m \omega }(x,x') ] = -\delta(x-x'),
\end{equation}
where 
\begin{equation}
\mathcal L:=x^{-2s}\frac{d}{dx} \left(x^{2s+2}\frac{d}{dx}\right) + V(x), 
\end{equation}
 with $V$ given by
\begin{equation}\label{eq:U}
V(x) := (k+m)\Big(k+ (k+m) (x+1)^2+2 i s x \Big) + \frac{k^2}{4x^2} + \frac{k^2}{x} + \frac{k(m-is)}{x} - {}_s\lambda_{\ell m \omega}.
\end{equation}
Here, we have introduced a shifted 
frequency,
\begin{align}
\ob :=  \omega - m\Omega_H = \omega- m,
\end{align}
where $\Omega_H := 1/(2r_H)=1$ is the horizon frequency.
The value $k=0$ (i.e., $\omega=m$) corresponds to
the so-called superradiant bound frequency (in the literature, this frequency is also called horizon frequency or critical frequency; we use these  terms interchangeably to denote $k=0$). 

In this paper we will carry out an in-depth analysis of the transfer function $\tilde g_{\indmode}(x,x')$.
For that purpose, we first define some
homogeneous solutions
of the radial equation \eqref{eq:radial teuk. Eq}.

\subsection{Radial Teukolsky equation}\label{sec:rad teuk}

The homogeneous version of the radial equation \eqref{eq:radial teuk. Eq},
\begin{equation}\label{eq:rad teuk x}
\mathcal L[R_{\ell m \omega}]=0,
\end{equation}
where $R_{\ell m \omega}=R_{\ell m \omega}(x)$ is a radial function,
is the key equation for frequency-domain perturbations of extremal Kerr.  This second-order, linear ordinary differential equation has rank-1 irregular singular points at infinity ($r=\infty$,
i.e., $x=\infty$) and at the horizon ($r=1/2$, i.e., $x=0$).  
This classifies it as a doubly confluent Heun equation \cite{ronveaux1995heun}.
This is in contrast with the radial equation in subextremal Kerr, which instead possesses two regular singular points (at the Cauchy and event horizons) and only one irregular singular point (at $x=\infty$), thus classifying it as a confluent Heun equation.
The main purpose of this paper is to, first, develop analytic techniques for solving 
Eq.~\eqref{eq:rad teuk x}
and, second, use these techniques to obtain late-time tails  of perturbations of extreme black holes.

In applications, it is natural to consider four different solutions to \eqref{eq:rad teuk x}, which, following Leaver's notation \cite{Leaver:1986a}, we denote by $R^{(0)}_{\pm}$ and $R^{(\infty)}_{\pm}$. These are defined according to boundary conditions imposed at the event horizon $x = 0$ and infinity 
\footnote{
The asymptotics (\ref{eq:Rin bc}) and (\ref{eq:Rup bc}) are only true boundary
conditions for the ingoing and upgoing solutions (i.e., they
specify the solutions uniquely for a choice of transmission
coefficients) when $\text{Im}(\omega)\geq 0$. The reason is that, for
$\text{Im}(\omega)<0$, the transmitted waves become exponentially dominant solutions in those asymptotic regions: $e^{-i\omega \ln x}\gg e^{i\omega \ln x}$ as
$x\to 0^+$ for in, and $e^{i\omega (x+\ln x)}\gg e^{-i\omega (x+\ln x)}$ as $x\to \infty$ for up.
Exponentially subdominant solutions in those regions are not
unambiguously determined by the asymptotic expressions.
Values of these solutions in the lower half plane must therefore
be determined instead by analytic continuation from $\text{Im}(\omega)\geq 0$.}.
In our notation, ``$0/\infty$" refers to the horizon/infinity, while ``$+/-$" means a purely outgoing/incoming wave.  For example, $R^{(\infty)}_{+}$ corresponds to the solution with purely outgoing radiation at infinity, while $R^{(0)}_{+}$ corresponds to radiation entering the black hole. Based on these properties, and following a more standard notation in the literature, we shall also denote
$R^{(\infty)}_{+}$ by the upgoing radial solution $\Rup$ and
$R^{(0)}_{+}$ by the ingoing radial solution $\Rin$.
The two notations are interchangeable. Mathematically, the ``in" and ``up" functions satisfy the following boundary conditions: 
\begin{align}\label{eq:Rin bc}
\Rin:= R^{(0)}_{+}
\sim 
\left\{\begin{array}{l l}
\displaystyle
\Rintra\ e^{i\ob/(2x)}x^{-2s}e^{-i\omega\ln x},&  x\to 0^+, \\
\displaystyle
\mathcal{I}_{\text{in}}\ \frac{e^{-i\omega ( x+ \ln x) }}{x}+
\mathcal{R}_{\text{in}}\ \frac{e^{i\omega ( x+ \ln x) }}{x^{1+2s}},& x\to \infty,
\end{array}
\right.
\end{align}
and
\begin{equation}
\label{eq:Rup bc}
\Rup := R^{(\infty)}_+ 
\sim
\left\{\begin{array}{l l}
\displaystyle
\Rupref\ e^{i\ob/(2x)}x^{-2s}e^{-i\omega\ln x}
+\Rupinc\ e^{-i\ob/(2x)}e^{i\omega\ln x}
,&  x\to 0^+,
\\
\displaystyle
\Ruptra\ \frac{e^{i\omega ( x+ \ln x) }}{x^{1+2s}},& x\to \infty,
\end{array}
\right.
\end{equation}
where 
$\mathcal I_{\text{in/up}}$,
$\mathcal R_{\text{in/up}}$ and
$\mathcal T_{\text{in/up}}$ are,
respectively,  complex-valued incidence, reflection and transmission coefficients of the ingoing/upgoing solutions.

The other homogeneous radial solutions
$R^{(0)}_{-}$ and $R^{(\infty)}_{-}$ 
obey boundary conditions such that they are
purely outgoing from the horizon and purely
ingoing from infinity, respectively  (see
Eqs.~(\ref{eq:R0+- near eh}) and (\ref{eq:Rinf+- near inf})).
Also following standard notation in the literature, we shall denote $R^{(0)}_{-}$
by the outgoing radial solution $\Rout$.
It satisfies
\begin{equation}\label{eq:Rout large-r}
\Rout:= R^{(0)}_{-}
\sim 
\Routtra\ e^{-i\ob/(2x)}e^{i\omega\ln x},\quad  x\to 0^+,
\end{equation}
where $\Routtra$ is its  transmission coefficient.

Clearly, any pair of the above solutions, such as
the pair $\{R^{(0)}_{-},R^{(\infty)}_{-}\}$
or  $\{R^{(0)}_{+},R^{(\infty)}_{+}\}$,
forms a complete set of linearly independent solutions of the homogeneous radial equation.

It shall prove useful to also
normalize radial solutions and coefficients in a different way.
We adopt the notation of placing a hat over a radial function or coefficient to indicate that quantity
normalized via the corresponding transmission coefficient:
\begin{equation}\label{eq:hatted quantities}
\hat R^{\text{in/up/out}}_{\indmode}:=  \frac{R^{\text{in/up/out}}_{\indmode}}{\mathcal{T}_{\text{in/up/out}}},
\quad
\mathcal{\hat I}_{\text{in/up}}:=  \frac{\mathcal I_{\text{in/up}}}{\mathcal{T}_{\text{in/up}}},
\quad
\mathcal{\hat R}_{\text{in/up}}:=  \frac{\mathcal R_{\text{in/up}}}{\mathcal{T}_{\text{in/up}}}.
\end{equation}
In particular, it follows from Eq.~\eqref{eq:Rin bc}
that
\begin{align}\label{eq:Rinhat bc}
\Rinhat
&\sim 
 e^{i\ob/(2x)}x^{-2s}e^{-i\omega\ln x},&  x\to 0^+, \\
\Ruphat & \sim \frac{e^{i\omega ( x+ \ln x) }}{x^{1+2s}},& x\to \infty.\label{eq:Ruphat bc}
\end{align}
In both of the asymptotic expressions in \eqref{eq:Rinhat bc} and \eqref{eq:Ruphat bc}, applying  the 
transformation $\left\{m\to -m, \omega\to-\omega^*\right\}$
is equivalent to complex conjugating them.
Similarly, using the symmetries in Eq.~\eqref{eq:symms eigen},
it is easy to see that 
applying $\left\{m\to -m,\omega\to-\omega^*\right\}$ on the radial operator $\mathcal L$ in Eq.~\eqref{eq:rad teuk x} is also equivalent to
complex conjugating it.
It then follows that 
\begin{equation}\label{eq:symm hat radial slns}
\Rinhat(x)=\Rinhatccmmmocc(x),\quad
  \Ruphat(x)=\Ruphatccmmmocc(x).  
\end{equation}
Similarly,  $\mathcal{\hat I}_{\text{in/up}}$
and $\mathcal{\hat R}_{\text{in/up}}$ are all complex conjugated under $\left\{m\to -m,\omega\to-\omega^*\right\}$.
This is the main reason for choosing the normalization
as in the hatted radial quantities.

\subsection{Transfer function}

The method we adopt for constructing the transfer function 
involves a set of linearly independent homogeneous solutions of the radial differential equation \eqref{eq:rad teuk x}.
The radial solutions yielding the \emph{retarded} Green function
of the Teukolsky equation are the above
 in and up solutions, which correspond to solutions  $\Psi$  of the Teukolsky equation 
 having no radiation coming out of the white hole or from past null infinity, respectively. 
The transfer function $\glmo(x,x')$ corresponding to the retarded Green function
is thus given by
\begin{equation}\label{eq:radialgreenfunction}
\glmo(x,x') = - \frac{R^{(0)}_+(x_<)R^{(\infty)}_+(x_>)}{\W},
\end{equation}
where $x_<:= \text{min}(x,x')$, $x_>:= \text{max}(x,x')$,
 $\W$ is the constant scaled Wronskian, 
\begin{align}
\W :=
\Delta^{s+1}
W[R^{(0)}_+,R^{(\infty)}_+]
=2 i\omega \mathcal{I}_{\text{in}}\Ruptra,\label{eq:Wequal}
\end{align}
and 
\begin{equation}
\Delta  := (r-r_H)^2 = x^2.
\end{equation}
We use the notation
\begin{equation}
W[R_1,R_2]:=
R_1\frac{dR_2}{dx}-R_2\frac{dR_1}{dx}
\end{equation}
for the actual Wronskian,
where $R_1$ and $R_2$ are any two solutions of the homogeneous radial equation.
We may equivalently  express the transfer function $\glmo(x,x')$ in terms of the hatted quantities as
\begin{equation}\label{eq:radialgreenfunction}
\glmo(x,x') =- \frac{\Rinhat(x_<)\Ruphat(x_>)}{\What} ,
\end{equation}
where $\hat{\mathcal W}$ is the scaled Wronskian, 
\begin{align}
\What :=
\Delta^{s+1} W[\Rinhat,\Ruphat] 
= 2 i\omega \Rininchat= ik \Rupinchat \label{eq:Whatequal,extreme}.
\end{align}
In the next-to-last equality in Eq.~\eqref{eq:Whatequal,extreme} we have evaluated the radial solutions for $x\to \infty$ and
in its last equality we have evaluated them for $x\to 0^+$.
It readily follows from Eq.~\eqref{eq:symm hat radial slns} that $\What$ is complex conjugated under $\left\{m\to -m,\omega\to-\omega^*\right\}$\footnote{We note that there is a typographical error in Eq.~~(3.17) ~\cite{Casals:Longo:2018}: the minus sign should not be present in its right-hand side, with no consequences at all for any of the results in~\cite{Casals:Longo:2018}.}.

We now give other scaled Wronskian identities which will be useful for our later calculations.
From the asymptotics in  Eqs.~\eqref{eq:Rin bc}, \eqref{eq:Rup bc} and \eqref{eq:Rout large-r}, it is straightforward to find
\begin{equation}\label{eq:W[Routhat,Ruphat]}
\Delta^{s+1}W[\Routhat,\Ruphat]=-ik\Ruprefhat
\end{equation}
and
\begin{equation}\label{eq:W[Routhat,Rinhat]}
\Delta^{s+1}W\left[\Routhat,\Rinhat\right]=-ik.
\end{equation}
The last Wronskian identity that we give is
\begin{align}\label{eq:W[Rinhat,Ruphatcc]}
\Delta^{s+1} W\left[\Rinhat,\Delta^{-s}\left.\Ruphatcc\right|_{-s}\right]=
&-2 i\omega 
\Rinrefhat
= ik\left.\Ruprefhatcc\right|_{-s},
\end{align}
which is only valid for $\omega\in\mathbb{R}$.
We note that, for $\omega\in\mathbb{R}$, if $R$ is a solution of the radial Teukolsky equation for spin $s\neq 0$, then, because  the corresponding radial operator $\mathcal L$ is not self-adjoint,
$R^*$ is not a solution of the equation (for any spin), but $\Delta^{-s}\left.R^*\right|_{-s}$ is a solution of the equation for spin $s$.

From Eqs.~\eqref{eq:radialgreenfunction} and \eqref{eq:Whatequal,extreme}, it follows that the symmetry of Eq.~\eqref{eq:symm hat radial slns} carries over to the transfer function modes:
$\glmo(x,x')=\glmoccmmmocc(x,x')$. Applying $\left\{m\to -m,\omega\to -\omega^*\right\}$ to the exponential factors in the integrand in Eq.~\eqref{Eq:TDGF} is
also equivalent to complex conjugating them.
If the whole integrand in Eq.~\eqref{Eq:TDGF} transformed in this way,
then the $\ell$-modes of the retarded Green function $G(x^\mu,x^\mu{}')$, as well as $G(x^\mu,x^\mu{}')$ itself, would be real valued.
However, under the transformation
$\left\{m\to -m,\omega\to -\omega^*\right\}$, Eq.~\eqref{eq:symms eigenfunc} shows that the angular factor ${}_s\eigenSS_{\indmode}(\theta,\theta')$ not only becomes complex-conjugated but also undergoes 
$\left\{\theta\to \pi-\theta,\theta'\to \pi-\theta'\right\}$ (or,
equivalently, it undergoes $s\to -s$).
This means that $G(x^\mu,x^\mu{}')$ is generally not real valued
(although it is real valued on the equator $\forall s$ and everywhere for $s=0$):
complex-conjugating it is equivalent to
taking $\left\{\theta\to \pi-\theta,\theta'\to \pi-\theta'\right\}$.
Mathematically, the fact that $G(x^\mu,x^\mu{}')$ is not real valued
 for $s\neq 0$ can be traced back to the fact that the Teukolsky operator is not self-adjoint for $s\neq 0$.

\section{MST method}\label{sec:MST}

In this section we bring forth the MST machinery and use it
to derive practical analytic expressions for the
various physically relevant solutions to the homogeneous radial Teukolsky equation and associated scattering amplitudes. We end the section with an exploration of the asymptotic behavior  
of the series coefficients and renormalized angular momentum as $\omega\to\{0,m\}$. 

We note that most of the results in this section have been numerically validated in~\cite{MSc-Longo,Casals:Longo:2018}.
 The numerical validation performed in these references consists of checking the following: (i) that values from different expressions in this paper for the same quantity (such as $\What$ in Eq.~\eqref{eq:Whatequal,extreme}) agree with each other; (ii) that such values are consistent with 
 limiting
 values from the MST formalism in subextremal Kerr evaluated for near-extremal values of $a$; and (iii)
 that values of quasinormal modes (which are poles in the complex frequency plane of the transfer function) obtained using expressions here agree with tabulated values
 in~\cite{Richartz:2016}.

\subsection{Series representations for the radial solutions}\label{sec:radial series}

\subsubsection{Radial series}\label{sec:Rseries}

Inspired by the work of Leaver \cite{Leaver:1986a} and MST
\cite{Mano:Suzuki:Takasugi:1996,Mano:1996mf,Sasaki:2003xr}, we make an ansatz for the homogeneous radial solutions as a sum over irregular confluent hypergeometric functions $U$. Our choices, which
are based on Eqs.~(191) and (192) of Ref.~\cite{Leaver:1986a}
(see Appendix ~\ref{sec:App radial Leaver} for further justification), are
\begin{align} \label{eq:MST up confl}
R^{(\infty)}_{\pm}&=\zeta^{(\infty)}_{\pm}x^{-s+\nu}e^{ik /(2x)}e^{\pm i\omega x}(2\omega)^{\nu+1}e^{- i \pi \e/2}e^{\mp i\pi(\nu+1/2)} \nn
 \\ &
 \times\sum_{n=-\infty}^{\infty} \left(\frac{\Gamma(\q{n}{\nu} + \e)}{\Gamma(\q{n}{\nu} -\e)}\right)^{1/2}   
  \left(\frac{\Gamma(\q{n}{\nu} \pm \e)}{\Gamma(\q{n}{\nu}  \mp \e )}\right)^{1/2}(-2i \omega x)^n  \aMST{n}
U\left(\q{n}{\nu} \pm  \e, 2 \q{n}{\nu},\mp 2i\omega x\right)
\end{align}
and
\begin{align} \label{eq:MST in confl}
R^{(0)}_{\pm}= &\zeta^{(0)}_{\pm}x^{-s-\nu-1}e^{i\omega x}e^{\pm ik/(2x)}k^{\nu+1}e^{- i \pi \et /2}e^{\mp i\pi(\nu+1/2)}
 \nonumber \\ & \times
\sum_{n=-\infty}^{\infty}
 \left(\frac{\Gamma(\q{n}{\nu} - \et )}{\Gamma(\q{n}{\nu} + \et )}\right)^{1/2}
 \left(\frac{\Gamma( \q{n}{\nu} \pm \et )}{\Gamma(\q{n}{\nu} \mp \et)}\right)^{1/2}
 \frac{\Gamma(\q{n}{\nu} + \e)}{\Gamma(\q{n}{\nu} -\e)}
\left(\frac{-i k}{x}\right)^n  \aMST{n}
U\left(\q{n}{\nu} \pm \et,2 \q{n}{\nu},\mp \frac{i k} {x}\right),
\end{align}
where $a_n^\nu$ are series coefficients. 
Here we have defined 
\begin{align} \label{eq:chi}
\e := s-i\omega,
\quad \et = -s-i\omega,
\end{align} 
as well as
\begin{align}
\q{n}{\nu} := n+\nu+1.
\end{align}
We have also introduced an auxilliary parameter $\nu$, the so-called renormalized angular momentum, which plays an important role in all MST analyses. 
The normalization constants $\zeta^{(0)}_{\pm}$ and $\zeta^{(\infty)}_{\pm}$ will be chosen in Sec.~\ref{sec:symm dec} such that the radial solutions are symmetric under $\nu\to-1-\nu$.
We have directly checked that Eqs.~\eqref{eq:MST up confl} and \eqref{eq:MST in confl} satisfy the homogeneous radial equation
\eqref{eq:rad teuk x} as long as the  coefficients $\aMST{n}$ satisfy a certain recurrence relation (see Eq.~\eqref{eq:rec rln an} below).
We shall deal with the series coefficients $\aMST{n}$ and with
$\nu$ in the next subsubsection.

 It follows from Eqs.~\eqref{eq:MST up confl} and \eqref{eq:MST in confl}, together with the analytical properties of the irregular confluent hypergeometric function $U$ (as well as the prefactors $\omega^{\nu+1}$ and $k^{\nu+1}$, respectively), that, in principle, $\omega=0$ is a branch point of the solutions $R^{(\infty)}_{\pm}$ and $\omega=m$ (i.e., $k=0$) is a branch point of the solutions 
 $R^{(0)}_{\pm}$.
 As we shall see in Secs.\ref{sec:BC origin} and \ref{sec:BC k=0}, these are indeed branch points of the radial solutions and they carry over to the transfer function.
 
We note that, in subextremal Kerr, while the corresponding $R^{(\infty)}_{\pm}$ solutions are similarly expressed in terms of the irregular confluent hypergeometric $U$ functions, the corresponding $R^{(0)}_{\pm}$ solutions are instead expressed in terms of the regular hypergeometric ${}_2F_1$ functions. This is due to the aforementioned fact that the event horizon is a regular singular point of the radial equation in subextremal Kerr whereas it is an irregular singular point in extremal Kerr.
As a consequence, the transfer function in subextremal Kerr only possesses a branch point at the origin, $\omega=0$,
which is responsible for the late-time decay of the linear perturbations~\cite{Leaver:1986,PhysRevLett.84.10,casals2016high}.
The branch point at $\omega=m$ is a new feature of the extremal configuration and gives rise to the Aretakis phenomenon of field perturbations on the horizon hole~\cite{Casals:2016mel, Gralla:2017lto}.

To simplify future calculations, we now give slightly more compact expressions for the radial solutions separately and obtain their transmission coefficients. For the ingoing and upgoing solutions, which are the ones of main interest here, the above expressions
simplify a little further:
\begin{align} \label{eq:Rup}
\Rup&=
f_{\rm up}(x,\omega)
\sum_{n=-\infty}^{\infty}
A^{up}_n(x,\omega)
U(\q{n}{\nu} + \e, 2 \q{n}{\nu}, -2 i \omega x)
\end{align}
and
\begin{align} \label{eq:Rin}
\Rin =
f_{\rm in}(x,k)
\sum_{n=-\infty}^{\infty} 
A^{in}_n(x,k)
U\left(\q{n}{\nu} + \et, 2 \q{n}{\nu}, - \frac{ik}{x} \right),
\end{align}
where 
\begin{subequations}
\begin{align}
&
f_{\rm up}(x,\omega):= \zetainfp x^{-s+\nu} e^{ik/(2x)}e^{ i\omega x}(2\omega)^{\nu+1}
e^{- i \e \pi /2} e^{- i \pi(\nu + \frac12)},
\\ &
f_{\rm in}(x,k):= \zetahorp x^{-s-\nu-1}e^{i\omega x} e^{i k /(2x)}k^{\nu +1 } e^{-i \pi \et /2} e^{- i \pi (\nu + \frac12)},
\end{align}
\end{subequations}
and
\begin{subequations}
\begin{align}
&
A^{up}_n(x,\omega):= \frac{\Gamma(\q{n}{\nu} + \e)}{\Gamma(\q{n}{\nu} - \e)}  (-2 i\omega x)^n a_n^\nu,
\\ &
A^{in}_n(x,k):= \frac{\Gamma(\q{n}{\nu} + \e )}{\Gamma(\q{n}{\nu} - \e)}\left(\frac{-ik}{x}\right)^n \aMST{n}.
\end{align}
\end{subequations}
The reason for writing $\omega$ as the argument of $f_{\rm up}$ but $k$ as that of $f_{\rm in}$ is to make manifest their branch points at $\omega=0$ and $k=0$ respectively
(we write the arguments of $A^{in}_n$ and $A^{up}_n$ merely out of notational consistency, not to denote any branch points in these functions).
The transmission coefficients, defined via Eqs.~(\ref{eq:Rin bc}) and (\ref{eq:Rup bc}),
readily follow from Eqs.~(\ref{eq:Rup}) and
(\ref{eq:Rin}). Using Eq.~(13.2.6)~\cite{NIST:DLMF}, we obtain
\begin{equation}\label{eq:Rintra}
\Rintra
 =\zetahorp k^{\nu +1 }(-ik)^{-\nu-1+s+i\omega} e^{-i \pi \et /2} e^{- i \pi (\nu + \frac12)} \sum_{n=-\infty}^{\infty} \frac{\Gamma(\q{n}{\nu} + \e )}{\Gamma(\q{n}{\nu} - \e)} \aMST{n}, 
\end{equation}
and
\begin{equation}\label{eq:Ruptra}
\Ruptra=
\zetainfp
e^{- i \e \pi /2} e^{- i \pi(\nu + \frac12)}(-2i\omega)^{-\nu-1-s+i\omega}(2\omega)^{\nu+1}
\sum_{n=-\infty}^{\infty}\frac{\Gamma(\q{n}{\nu} + \e)}{\Gamma(\q{n}{\nu} - \e)}  a_n^\nu. 
\end{equation}

It will also be useful to give explicit expressions for $R^{(0)}_{-}\equiv\Rout$ and its transmission coefficient near the horizon.
From Eq.~\eqref{eq:MST in confl} we find that
\begin{equation} \label{eq:Rout}
\Rout =
f_{\rm out}(x,k)
\sum_{n=-\infty}^{\infty} 
A^{out}_n(x,k)
U\left(\q{n}{\nu} - \et, 2 \q{n}{\nu},  \frac{ik}{x} \right),
\end{equation}
where 
\begin{equation}
f_{\rm out}(x,k):= \zetahorm x^{-s-\nu-1}e^{i\omega x} e^{-i k /(2x)}k^{\nu +1 } e^{-i \pi \et /2} e^{ i \pi (\nu + \frac12)}
\end{equation}
and
\begin{equation}
A^{out}_n(x,k):= \left(\frac{-ik}{x}\right)^n \aMST{n}.
\end{equation}
The coefficient of the outgoing solution near the horizon is readily obtained from Eq.~\eqref{eq:Rout} and
Eq.~(13.2.6)~\cite{NIST:DLMF}:
\begin{equation}\label{eq:Routtra}
\Routtra
 =\zetahorm k^{\nu +1 }(ik)^{-\nu-1-s-i\omega} e^{-i \pi \et /2} e^{ i \pi (\nu + \frac12)} \sum_{n=-\infty}^{\infty}(-1)^n \aMST{n}.
\end{equation}


\subsubsection{Series coefficients and renormalized angular momentum}\label{sec:coeffs}

%
%
%
%
%
%
%
%
%
%
%

Here we give recurrence relations for the MST coefficients $\aMST{n}$ and discuss the properties of the solutions of these relations.

Using\footnote{Equations \eqref{eq:rec rln U}
and \eqref{eq:rec rln dU} are given below Eq.~(3.15) in~\cite{Casals:Ottewill:2015} but here we correct a typographical error there of an extra factor of $i$ in front of
$\hat{H}^+_{L+1}(-\eta,z)$ and of $\hat{H}^+_{L-1}(-\eta,z)$.}
\begin{align}\label{eq:rec rln U}
&\frac{1}{z} \,
  \hat{H}^+_L(-\eta,z) =-i\frac{(L+1-i \eta) }{(L+1) (2L+1)}
   \hat{H}^+_{L+1}(-\eta,z)+\frac{\eta }{L(L+1)} \hat{H}^+_{L}(-\eta,z)+i\frac{(L+i \eta) }{L (2L+1)}
    \hat{H}^+_{L-1}(-\eta,z),\\
& \frac{\text{d}\ }{\text{d}z} \,
   \hat{H}^+_L(-\eta,z) = i\frac{L(L+1-i \eta) }{(L+1) (2L+1)}
   \hat{H}^+_{L+1}(-\eta,z)+\frac{\eta }{L(L+1)} \hat{H}^+_{L}(-\eta,z)+\frac{(L+1)(L+i \eta) }{L (2L+1)}
   \hat{H}^+_{L-1}(-\eta,z),
   \label{eq:rec rln dU}
\end{align}
where 
\begin{align}
\hat{H}^+_L(\eta,z)  :=  e^{i z} (-2i z)^{L+1} U(L+1+i\eta,2L+2,-2iz) ,
\end{align}
we find that 
Eqs.~\eqref{eq:MST up confl} and \eqref{eq:MST in confl} satisfy 
Eq.~\eqref{eq:rad teuk x} as long as 
the
coefficients $\aMST{n}$ satisfy the following bilateral recurrence relation: 
\begin{equation}\label{eq:rec rln an}
\alpha_n \aMST{n+1}+
\beta_n\aMST{n}+
\gamma_n \aMST{n-1}=0,\quad n\in \mathbb{Z},
\end{equation}
where
\begin{align}
\alpha_n &:= \frac{
\eps
(\q{n}{\nu}+\e)(\q{n}{\nu} - \et)}{\q{n}{ \nu} (2 \q{n}{ \nu} +1 )},\nonumber\\
\beta_n &:= ( \q{n}{ \nu} -1 ) \q{n}{\nu} -
\eigenbA
- \epsilon\,  \frac{ \e \et}{ (\q{n}{ \nu} -1 ) \q{n}{ \nu}  }, \label{eq:rec.Eq. coeffs} \\ 
\gamma_n &:= \frac{
\eps
(\q{n}{ \nu} -1 -\e)(\q{n}{ \nu} -1 + \et)}{(\q{n}{\nu}-1)(2\q{n}{ \nu} -3)},
\nonumber
\end{align}
and
$\eigenbA:= -\frac74 \omega^2 +s(s+1) + \eigenA$.
Here we have defined \[ \eps:= \omega k .\]
We  note that, under $\nu\to -\nu-1$, $\alpha_n$ transforms to $\gamma_{-n}$ and $\beta_n$ transforms to $\beta_{-n}$.
We choose the normalization $\aMSTmnu{0}=\aMST{0}=1$ and, therefore, $\aMSTmnu{n}=\aMST{-n}$, $\forall n\in \mathbb{Z}$, directly follows.

%
%
%
%
%
%
%
%
%
%
%
%
%
%
%

The bilateral 
recurrence relations Eq.~(\ref{eq:rec rln an}) may be solved in the following way.
First, define the ratios
\begin{equation}
R_n:= \frac{\aMST{n}}{\aMST{n-1}},\quad
L_n:= \frac{\aMST{n}}{\aMST{n+1}}.
\end{equation}
Then, using the recurrence relations, express these ratios as the following continued fractions:
\begin{align}\label{eq:cont fracs}
&
R_n=-\frac{\gamma_n}{\beta_n+\alpha_n R_{n+1}}=-\frac{\gamma_n}{\beta_n-}\cdot\frac{\alpha_n\gamma_{n+1}}{\beta_{n+1}-}\cdot\frac{\alpha_{n+1}\gamma_{n+2}}{\beta_{n+2}-}\dots
\\ &
L_n=-\frac{\alpha_n}{\beta_n+\gamma_n L_{n-1}}=-\frac{\alpha_n}{\beta_n-}\cdot\frac{\alpha_{n-1}\gamma_{n}}{\beta_{n-1}-}\cdot\frac{\alpha_{n-2}\gamma_{n-1}}{\beta_{n-2}-}\dots .
\label{eq:cont fracs Ln}
\end{align}
Now, once $R_n$ and $L_n$ have been obtained (either numerically to within a certain prescribed precision or analytically up to a certain order in an expansion parameter),
respectively, $\forall n>0$ and  $\forall n<0$, then one can obtain $\aMST{n}=R_n \aMST{n-1}$,
$\forall n>0$, by starting from $n=1$ 
(given a certain choice for $\aMST{0}$ as a  normalization choice, such as ours, $\aMST{0}=1$).
Similarly, one can obtain $\aMST{n}=L_n \aMST{n+1}$, $\forall n<0$.

To investigate the convergence of the continued fractions we carry out the large-$n$ asymptotics of the series coefficients $\aMST{n}$.
In order to find the large-$n$ behavior of the solutions 
of the recurrence relations Eq.~\eqref{eq:rec rln an},
we apply Theorem 2.3 of Ref.~\cite{gautschi1967computational}.
We find that there exists a pair of solutions, say $a_n^{(1)}$ and $a_n^{(2)}$, which satisfy
\begin{align}\label{eq:a_n large-n}
\frac{a_{n+1}^{(1)}}{a_n^{(1)}}\sim -\frac{
\eps
}{2}\frac{1}{n^2},
\quad
\frac{a_{n+1}^{(2)}}{a_n^{(2)}}\sim -\frac{2}{
\eps
}n^2,
\quad n\to+\infty,
\end{align}
and another pair,
say $b_n^{(1)}$ and $b_n^{(2)}$, which satisfy
\begin{align}\label{eq:b_n large-n}
\left|\frac{b_{n}^{(1)}}{b_{n+1}^{(1)}}\right|
\sim
\frac{
\left|\eps\right|
}{2}\frac{1}{n^2},
\quad
\left|\frac{b_{n}^{(2)}}{b_{n+1}^{(2)}}\right|
\sim \frac{2}{
\left|\eps\right|
}n^2,
\quad n\to-\infty.
\end{align}
Since $\lim_{n\to + \infty}a_n^{(1)}/a_n^{(2)}=0$, $a_n^{(1)}$ is said to be a {\it minimal} solution (which is unique up to a normalization) and $a_n^{(2)}$ a {\it dominant} solution as $n\to +\infty$; similarly, $b_n^{(1)}$ is said to be a minimal solution  and $b_n^{(2)}$ a dominant solution as $n\to -\infty$.
The continued fraction in Eq.~\eqref{eq:cont fracs} [resp. Eq.~\eqref{eq:cont fracs Ln}]
converges when applied to a minimal solution
as $n\to +\infty$ [$n\to -\infty$]
(see Theorem 1.1~ of Ref. \cite{gautschi1967computational}).
The renormalized angular momentum parameter $\nu$ is determined by the consistency requirement that
the 
minimal solution as $n\to \infty$ coincides with the minimal solution as $n\to -\infty$, i.e., that
$a_n^{(1)}=b_n^{(1)}$\MC{Here you had added ``as $n\to \pm \infty$" but I removed it as I don't see why it should be there}.
The series coefficients $\aMST{n}$ shall henceforth denote the corresponding unique minimal solution of the recurrence relations Eq.~(\ref{eq:rec rln an}) as $n\to \pm \infty$
 with the normalization choice $\aMST{0}=1$.
 
The choice of  $\aMST{n}$ as a minimal solution as both $n\to \infty$ and $n\to -\infty$
 guarantees~\cite{Leaver:1986a} that the series in
 Eq.~\eqref{eq:MST up confl}
 converges everywhere except on $r=M$  and that the series
in Eq.~\eqref{eq:MST in confl}
 converges everywhere except on $r=\infty$.
This means that the value of $\nu$ is fixed via the following condition:
\begin{align}\label{eq:eq for nu}
R_n L_{n-1} = 1.
\end{align}
Equivalently, one may impose the condition
\begin{align}\label{eq:eq for nu bis}
\beta_{n}+\alpha_n R_{n+1}+\gamma_n L_{n-1} = 0.
\end{align}
This is a transcendental equation for $\nu$ where $R_n$ and $L_{n}$ may be obtained from the continued fractions in
Eqs.~\eqref{eq:cont fracs}
and \eqref{eq:cont fracs Ln}.
One is free to choose the value of $n\in\mathbb{Z}$ in Eqs.~(\ref{eq:eq for nu}) and (\ref{eq:eq for nu bis}).

The recurrence relations Eq.~\eqref{eq:rec rln an} in extremal Kerr are the same as in subextremal Kerr [see, e.g., Eqs.~(123) and (124) in~\cite{Sasaki:2003xr}] when taking the extremal limit $a\to M$, except for a change in the signs of $\alpha_n$ and of $\gamma_n$.
Such sign changes simply amount to $\aMST{n}\to (-1)^n\aMST{n}$ and they do not affect Eq.~\eqref{eq:eq for nu}, which determines $\nu$.
Therefore, general properties  of $\nu$ in subextremal Kerr which have been derived in the literature from its defining equation are also satisfied by $\nu$ in extremal Kerr.
We next note some
properties and symmetries exhibited by the MST construction and which relate to the renormalized angular momentum parameter $\nu$\footnote{We note that the parameter $\nu$ appears in other analyses of black hole perturbations which do not use the MST formalism. See, for example,~\cite{castro2013black} in subextremal Kerr, where $\nu$ is related to the monodromy of the upgoing radial solution at the irregular singular point $r=\infty$. See also  Sec.\ref{sec:coeffs,nu omega->0,m} here.}: 

\begin{itemize}

\item The  MST formalism is fundamentally invariant under $\nu \to -\nu - 1$.  
The reason is that $\nu$ was introduced as a parameter in the radial ordinary differential equation through the combination ``$\nu(\nu+1)$''  [see Eq.~(119)~\cite{Sasaki:2003xr} in the subextremal case].
This leads, in particular, to the symmetry
\begin{equation}\label{eq:an->a-n}
\aMSTmnu{n}=\aMST{-n}
\end{equation}
observed previously. 
As long as we require the boundary conditions of the radial solutions to also be invariant under $\nu \to -\nu - 1$, the radial solutions will also satisfy this symmetry [see Eq.~\eqref{symmetric decomp}].

\item If $\nu$ is a solution of Eq.~(\ref{eq:eq for nu bis}), then so is $\nu + k$, for any $k\in \mathbb{Z}$.
The reason is that $\nu$ only appears in Eqs.~(\ref{eq:eq for nu}) and (\ref{eq:eq for nu bis}) in the combination $\nu+n$, where $n\in \mathbb{Z}$.

\item Applying  $\nu\to \nu^*$ and $\omega\to \omega^*$ to all coefficients $\alpha_{n}$, $\beta_{n}$ and $\gamma_{n}$ is equivalent,
from their definitions, to complex conjugating them
\footnote{For $\beta_n$, we use the  first two properties in Eq.~\eqref{eq:symms eigen}.}.
This implies, from Eq.~\eqref{eq:eq for nu bis},
that applying  $\omega\to \omega^*$ to $\nu$ is equivalent to complex conjugating $\nu$.
\MC{I don't have a proof that the following is the only exception}
We note that these transformation properties are, however, not necessarily true if $\omega$ lies on a branch cut of $\nu$; in that case, $\nu$ is not necessarily  real when $\omega$ is real.

\item It has been shown (analytically, but with an assumption which is supported numerically) in subextremal Kerr in~\cite{fujita2005new,Sasaki:2003xr} and in extremal Kerr in~\cite{MSc-Longo,Casals:Longo:2018} that, for $\omega$ real,
$\nu$ is either real valued or else complex valued with a real part that is equal to a half-integer number.

\item It follows from the above property that, for $\omega$ real, complex conjugation of $\nu$ can be achieved by applying the MST symmetries of $\nu \to -\nu - 1$ and the addition of an integer to $\nu$. 

\item The series coefficients
$\alpha_{n}$,
$\beta_{n}$ and $\gamma_{n}$
are all invariant\footnote{For $\beta_n$,
we use the last property in Eq.~\eqref{eq:symms eigen}.}
under $m\to -m$ and
$\omega\to -\omega$, and, therefore, so is $\nu$.
\end{itemize}

We note that throughout the paper we make use of the symmetry \eqref{eq:an->a-n} in the $n$-sums of the MST series.

\subsection{Symmetric decomposition of the radial solutions}\label{sec:symm dec}

We now write the radial solutions in a form that is manifestly invariant under $\nu\to -\nu-1$.
For that purpose, we use Eq.~(13.2.42)~\cite{NIST:DLMF} to write the solutions in terms of the {\it regular} confluent hypergeometric function $M$ as\footnote{We note that here we choose to use the subindex ``$\infty$" for the solutions which, in the subextremal case, correspond to those in~\cite{Sasaki:2003xr} with subindex ``$C$", referring to Coulomb wave functions.
The reason is that in the extremal case both the ingoing and upgoing solutions have representations in terms of Coulomb wave functions.}
\begin{subequations}\label{symmetric decomp}
\begin{align}
&\label{Rranger}
R^{(\infty)}_{\pm}=R^{\nu}_{\Cinf,\pm}+R^{-\nu-1}_{\Cinf,\pm},
\\
&\label{R0pm}
R^{(0)}_{\pm}=R^{\nu}_{0,\pm}+R^{-\nu-1}_{0,\pm},
\end{align}
\end{subequations}
where
\begin{align}
R^{\nu}_{\Cinf,\pm}
:= &\,\,\,\zeta^{(\infty)}_{\pm}x^{-s+\nu}e^{ik/(2x)}e^{\pm i\omega x}(2\omega)^{\nu+1}e^{-\pi i\e/2}e^{\mp i\pi(\nu+1/2)}
 \\ & \times \sum_{n=-\infty}^{\infty}
\left(\frac{\Gamma(\q{n}{\nu}+\e)}{\Gamma(\q{n}{\nu}-\e)}\right)^\frac12
 \left(\frac{\Gamma(\q{n}{\nu}\pm\e)}{\Gamma(\q{n}{\nu}\mp \e)}\right)^\frac12
\frac{\Gamma(1-2 \q{n}{\nu})}{\Gamma(1-\q{n}{\nu}\pm \e)}(-2i\omega x)^n  \aMST{n}M\left(\q{n}{\nu}\pm \e,2\q{n}{\nu},\mp 2i\omega x\right)
\nonumber
\end{align}
and 
{\small
\begin{align}\label{eq:R^-nu-1_0,pm}
&R^{-\nu-1}_{0,\pm} := \,\,\,
\zeta^{(0)}_{\pm}\vert_{\nu}\,\, x^{-s-\nu-1}e^{i\omega x}e^{\pm ik/(2x)}k^{\nu+1}e^{- i \pi \et /2}e^{\mp i\pi(\nu+1/2)}    \\ & \times
\sum_{n=-\infty}^{\infty}
 \left(\frac{\Gamma(\q{n}{\nu}- \et )}{\Gamma(\q{n}{\nu} + \et)}\right)^\frac12
 \left(\frac{\Gamma(\q{n}{\nu} \pm \et)}{\Gamma(\q{n}{\nu} \mp \et)}\right)^\frac12
 \frac{\Gamma(\q{n}{\nu}  + \e )}{\Gamma(\q{n}{\nu}  - \e)}
  \frac{\Gamma(1-2\q{n}{\nu})}{\Gamma(1-\q{n}{\nu} \pm \et)}
\left(-\frac{ik}{x}\right)^n  \aMST{n}
M\left(\q{n}{\nu} \pm \et,2\q{n}{\nu},\mp \frac{ik}{x}\right).
\nonumber 
\end{align}}
The decompositions \eqref{Rranger} and \eqref{R0pm} simplify for the in and up solutions. For these, we find
\begin{subequations}
\begin{align}\label{eq:RnuC+}
R^{\nu}_{\Cinf,+}  = &\,\,\, \zetainfp 
x^{-s+\nu}e^{ik/(2x)}e^{ i\omega x}(2\omega)^{\nu+1}e^{-i \pi \e /2}e^{- i\pi(\nu+\frac12)}\nonumber\\
&\times
\sum_{n=-\infty}^{\infty} \frac{ \Gamma(\q{n}{\nu}+\e) \Gamma(1-2 \q{n}{\nu} )  }{ \Gamma(\q{n}{\nu}- \e) \Gamma(1-\q{n}{\nu} +\e)} \left(-2 i \omega x\right)^n \aMST{n} M\left(\q{n}{\nu} +\e , 2 \q{n}{\nu} , -2 i \omega x \right),\\ 
 R^{-\nu -1}_{\Cinf,+} = &\,\,\,\zetainfp
 x^{- \nu -1 -s} e^{ik/(2x)} e^{ i \omega x}(2 \omega)^{\nu+1}(-2 i \omega)^{-2\nu-1} e^{- i \pi \e/2}e^{-i\pi(\nu+\frac12)}
\nonumber \\ 
&\times 
\sum_{n=-\infty}^{\infty} \frac{ \Gamma(2 \q{n}{\nu}-1) }{\Gamma(\q{n}{\nu} -\e) } (-2 i \omega x)^{-n} \aMST{n} M\left(1 - \q{n}{\nu} + \e, 2(1-\q{n}{\nu}\right), - 2 i \omega x),
\label{eq:R-nu-1C+}
\end{align}
\end{subequations}
and
\begin{subequations}
\begin{align}
R^{\nu}_{0,+}=& \,\,\,\zetahorp 
x^{-s+\nu}e^{ik/(2x)}e^{ i\omega x}k^{\nu+1}(-ik)^{-2\nu-1}e^{- i \pi \et/2} e^{-i\pi(\nu+\frac12)}
\nonumber \\
&\times \sum_{n=-\infty}^{\infty} \frac{ \Gamma(\q{n}{\nu} + \e) \Gamma( 2 \q{n}{\nu}-1) }{\Gamma(\q{n}{\nu}- \e)\Gamma(\q{n}{\nu}+ \et ) } \aMST{n} \left( \frac{ix}{k} \right)^{n} M\left(1-\q{n}{\nu} + \et,2(1-\q{n}{\nu}), -\frac{ik}{x} \right), 
\label{eq:R_0+^nu}
\\ 
R^{-\nu-1}_{0,+} = & \,\,\,\zetahorp 
x^{-s-\nu-1}e^{i\omega x}e^{ ik/(2x)}k^{\nu+1}e^{- i \pi\et /2}e^{- i\pi(\nu+1/2)}\nonumber \\ &\times \sum_{n=-\infty}^{\infty} \frac{ \Gamma(\q{n}{\nu} + \e) \Gamma(1- 2 \q{n}{\nu}) }{\Gamma(\q{n}{\nu}- \e) \Gamma(1-\q{n}{\nu}+\et)} \left(  \frac{ k}{ix} \right)^n \aMST{n} M \left(\q{n}{\nu}+ \et, 2 \q{n}{\nu}, - \frac{ik}{x} \right) .
\label{eq:Rnu-1_0,+}
\end{align}
\end{subequations} 
A transformation property we have used here is
\begin{align}
\frac{ \Gamma(\q{n}{\nu} + \e)}{\Gamma(\q{n}{\nu}-\e)} \to 
\frac{\sin(\pi(\nu-i\omega))}{\sin(\pi(\nu+i\omega))}
\frac{ \Gamma(\q{n}{\nu} + \e)}{\Gamma(\q{n}{\nu}-\e)} , \quad \text{under}\quad \nu \to -\nu-1,
\end{align}
for $s\in \mathbb{Z}$.

Equations (\ref{Rranger}) and (\ref{R0pm}) put into manifest the symmetry of the radial solutions under $\nu\to -\nu-1$ provided the $\nu$-dependent normalizations are chosen appropriately.
In particular, for Eqs.~\eqref{eq:RnuC+}--\eqref{eq:Rnu-1_0,+} to respect the symmetry under $\nu\to -\nu-1$ in Eqs.~\eqref{symmetric decomp} we  require
\begin{subequations}
\begin{align}
    \label{eq:transf zetas0}
  \zeta^{(0)}_+
  &\to
  e^{-2 i \pi (\nu+\frac12 )}
    (-ik)^{-2\nu-1}k^{2\nu+1}
    \frac{\sin \left( \pi ( \nu + i \omega ) \right)}{ \sin \left( \pi (\nu- i \omega) \right)}
    \zeta^{(0)}_+
    , \quad\ \ \ \text{under}\quad \nu \to -\nu-1,
    \\
    \zeta^{(\infty)}_+
    &\to
    e^{- 2 \pi i (\nu+\frac12 )}\left(-i\omega\right)^{-2\nu-1}\omega^{2\nu+1}  \frac{\sin \left( \pi ( \nu + i \omega ) \right)}{ \sin \left( \pi (\nu- i \omega) \right)}
    \zeta^{(\infty)}_+
    , \quad \text{under}\quad \nu \to -\nu-1.
\label{eq:transf zetas inf}
\end{align}
\label{eq:transf zetas}
\end{subequations}
The relations \eqref{eq:transf zetas} are satisfied by  
\begin{subequations}\label{eqs:zeta relations}
\begin{align}\label{eq: zeta 0}
 \zeta_{+}^{(0)}
 &= k^{-\nu}\left(-ik\right)^{\nu}e^{ i\pi \nu}  \left(\frac{\sin (\pi(\nu-i \omega))}{\sin (\pi (\nu+ i \omega))}\right)^{1/2},
   \\
    \label{eq: zeta inf}
    \zeta_{+}^{(\infty)}
    &= 
    \omega^{-\nu}\left(-i\omega\right)^{\nu}e^{ i\pi \nu}  \left(\frac{\sin (\pi(\nu-i \omega))}{\sin (\pi (\nu+ i \omega))}\right)^{1/2},
\end{align}
\end{subequations}
thereby fixing our normalization of the radial solutions.

\subsection{Matching radial solutions}\label{sec:matching}

We now match the radial solutions.  As both the in and up series solutions 
appearing in the right-hand sides of Eq.~\eqref{symmetric decomp}
converge in $M<r<\infty$, we proceed similarly to Sec.~4.4~\cite{Sasaki:2003xr} and match within the large overlap region of convergence. 
The $\nu \to -\nu -1$ symmetry of the radial solutions halves the amount of work necessary to match the solutions. Consequently, we choose to explicitly match $R^{\nu}_{\Cinf,+}$ to $R^{\nu}_{0,+}$ and obtain the  $R^{-\nu-1}_{\Cinf,+}$ to $R^{-\nu-1}_{0,+}$ by symmetry.  

Now, we write the hypergeometric functions appearing in  $R^{\nu}_{\Cinf,+}$ and $R^{\nu}_{0,+}$ as power series in $x$ and match the coefficients. From Eq.~(\ref{eq:RnuC+}) and Eq.~(13.2.2) of~\cite{NIST:DLMF} we readily obtain
\begin{equation}\label{eq:RnuC+ exp}
R^{\nu}_{\Cinf,+}=
 \zetainfp x^{-s+\nu}e^{ik/(2x)}e^{ i\omega x}(2\omega)^{\nu+1}e^{- i \pi \e /2}e^{- i\pi(\nu+1/2)}
\sum_{p=-\infty}^{\infty}
\left(\sum_{n=-\infty}^p D_{n,p-n}\right) x^p,
\end{equation}
where
\begin{equation}\label{eq:Dnj}
D_{n,j}:=  \frac{\Gamma(\q{n}{\nu} + \e )\Gamma(1-2 \q{n}{\nu})}{\Gamma(\q{n}{\nu}- \e)\Gamma(1-\q{n}{\nu}+\e)}
\frac{(\q{n}{\nu} + \e)_j}{(2 \q{n}{\nu})_j\ j!} \aMST{n}(-2 i\omega)^{n+j},
\end{equation}
and where $(z)_n:=\Gamma(z+n)/\Gamma(z)$ denotes the Pochhammer symbol. 
Similarly,  we obtain
\begin{equation} \label{eq:Rnu0+ exp}
R^{\nu}_{0,+}=
\zetahorp x^{-s+\nu}e^{ik/(2x)}e^{ i\omega x}k^{\nu+1}(-ik)^{-2\nu-1}e^{-i\pi/2}e^{- i \pi \et/2} e^{-i\pi\nu}
\sum_{p=-\infty}^{\infty}
\left(\sum_{n=p}^{\infty} C_{n,n-p}\right) x^p,
\end{equation}
where
\begin{equation}\label{eq:Cnj}
C_{n,j}:=
\frac{\Gamma(\q{n}{\nu} + \e)}{\Gamma(\q{n}{\nu}-\e)}
\frac{\Gamma(2\q{n}{\nu}-1)}{\Gamma(\q{n}{\nu} + \et)}
\frac{(1-\q{n}{\nu}+ \et)_j}{(2-2\q{n}{\nu})_j 
\ j!
} \aMST{n}(-ik)^{j-n}. 
%
\end{equation}

By comparing Eqs.~(\ref{eq:RnuC+ exp}) and (\ref{eq:Rnu0+ exp}) we see that they are proportional:
\begin{equation}\label{eq:Rnu0+ vs RnuC+}
R^{\nu}_{0,+}=K_{\nu}R^{\nu}_{\Cinf,+},
\end{equation}
with
\begin{equation}\label{eq:Knu}
K_{\nu}:= \frac{ \zetahorp }{ \zetainfp}k^{\nu+1}(-ik)^{-2\nu-1}(2\omega)^{-\nu-1}e^{i \pi s}\frac{\sum_{n=\rArbit}^{\infty} C_{n,n-\rArbit}}{\sum_{n=-\infty}^\rArbit D_{n,\rArbit-n}},
%
\end{equation}
and $\rArbit$ is an arbitrary integer.
From Eqs.~\eqref{eq:Rnu0+ vs RnuC+} and~\eqref{R0pm} it follows that
\begin{equation}\label{eq:Rin - RC}
\Rin =K_{\nu}R^{\nu}_{\Cinf,+}+K_{-\nu-1}R^{-\nu-1}_{\Cinf,+}.
\end{equation}
This series representation of the in solution is valid at $r=\infty$ and will be used in the next section to find the ingoing radial coefficients at infinity.

Finally, a representation for the up solution which is valid 
in the horizon limit
follows trivially from Eqs.~(\ref{Rranger}) and (\ref{eq:Rnu0+ vs RnuC+}):
\begin{equation}\label{eq:Rup from Rin}
\Rup=\left(K_{\nu}\right)^{-1}R^{\nu}_{0,+}+\left(K_{-\nu-1}\right)^{-1}R^{-\nu-1}_{0,+}.
\end{equation}
This series representation
will be used in a later section to find the upgoing radial coefficients at the horizon.


\subsection{Radial coefficients (scattering amplitudes)}
\label{sec:ampl}

In Sec.\ref{sec:symm dec} we obtained series representations for the transmission coefficients of the in, up and out radial solutions. In this subsection we derive series representations for the remaining radial coefficients: the incidence and reflection amplitudes. 


\subsubsection{Scattering amplitudes at infinity}\label{sec:ampl inf}
In order to obtain the radial coefficients at radial infinity, we
split $R^{\nu}_{\Cinf,+}$ into two pieces: one which
is purely ingoing and the other one purely outgoing at infinity.
We note that these ingoing and outgoing pieces are, of course, proportional to, respectively, 
$R^{(\infty)}_{-}$ and $R^{(\infty)}_{+}$.
For notation compactness, we label these new  ingoing and outgoing solutions with a new variable: $R_+^{\nu}$ and $R_-^{\nu}$, respectively.
Specifically, using Eq.~(6.7.7) Vol.1~\cite{Erdelyi:1953}, we split $R^{\nu}_{\Cinf,+}$ in Eq.~\eqref{eq:RnuC+} 
as 
\begin{equation}\label{eq:RC - R+/-}
R^{\nu}_{\Cinf,+}=R_+^{\nu}+R_-^{\nu},
\end{equation}
where
\begin{align}\label{eq:R_+^nu}
\Rpnu :=  &\,\, \zetainfp x^{-s+\nu}e^{ik /(2x)}e^{-i\omega x}(2\omega)^{\nu+1}e^{-\pi \omega(1\pm 2)/2}e^{- i\pi\nu(1\mp 1)-\pi i (s+1)/2}
 \\ &\times \frac{\sin(\pi(\nu+i\omega))}{\sin(2\pi\nu)}\sum_{n=-\infty}^{\infty}
( -2 i \omega x)^n  \aMST{n}U\left(\q{n}{\nu} - \e , 2\q{n}{\nu}, 2i\omega x\right),
\nonumber
\end{align}
and 
\begin{align}\label{eq:R_-^nu}
R^{\nu}_{-} := &\,\, \zetainfp x^{-s+\nu}e^{ik / (2x) }e^{ i\omega x}(2\omega)^{\nu+1}e^{-\pi \omega(1\pm 2)/2}e^{- i\pi\nu(1\pm 1)-\pi i (s+1)/2} \nonumber \\
&\times \frac{\sin(\pi(\nu+i\omega))}{\sin(2\pi\nu)}
\sum_{n=-\infty}^{\infty}
\frac{\Gamma(\q{n}{\nu} + \e )}{\Gamma(\q{n}{\nu} - \e)}
(- 2 i \omega x)^n  \aMST{n}U\left(\q{n}{\nu} + \e,2\q{n}{\nu},- 2i\omega x\right),
\end{align}
where the upper/lower signs, respectively, correspond to $\text{Re}(\omega x)>0/\text{Re}(\omega x)<0$\MC{what about $\text{Re}(\omega x)=0$?}. In deriving these relations  we have assumed $s\in\mathbb{Z}$.  By comparing Eq.~(\ref{eq:R_-^nu}) with Eq.~(\ref{eq:Rup}) we see that $R^{\nu}_{-}$
is  proportional to $R_{\rm up}:= R_+^{(\infty)}$
\footnote{Mathematically, this comes from the fact that in Eq.~\eqref{symmetric decomp} we split the irregular $U$ function appearing in the series for $\Rup$ into two regular $M$ functions; in Eq.~\eqref{eq:RC - R+/-}, in some sense we  ``undo'' this transformation by splitting each of these $M$ functions back into $U$ functions.}.
Similarly, comparing Eq.~(\ref{eq:R_+^nu}) with Eq.~(\ref{eq:MST up confl}), it follows that
$\Rpnu$ is 
proportional to
$R_-^{(\infty)}$.
Therefore, $R^{\nu}_{\pm}$  are homogeneous solutions of the Teukolsky equation.
It then follows from Eq.~(\ref{eq:RC - R+/-}) that $R^{\nu}_{\Cinf,+}$ is also a
homogeneous solution of the Teukolsky equation and from Eq.~(\ref{eq:Rnu0+ vs RnuC+}) that so is $R^{\nu}_{0,+}$.

In order to find the large-$x$ asymptotics of these solutions we use Eq.~(13.7.3)~\cite{NIST:DLMF}
to obtain
\begin{align}\label{eq:Rnu+ large-x}
&
\Rpnu\sim
\Rpnutra
\frac{e^{-i\omega \left( x+ \ln x \right)}}{x},
\quad x\to \infty,
\end{align}
and 
\begin{align}\label{eq:Rnu- large-x}
\Rmnu \sim  & \,\, 
\Rmnutra
\frac{e^{i\omega (x+  \ln x)}}{x^{1+2s}}, \quad x \to \infty,
\end{align}
for which
\begin{subequations}
\begin{align}\label{eq:Rnu+,tra}
\Rpnutra &:=
 \zetainfp 2^{s-i\omega}\omega^{\nu+s}(i\omega)^{-\nu-i\omega}
 e^{-\pi \omega(1\pm 2)/2}e^{- i\pi\nu(1\mp 1)}e^{-\pi i}
\frac{\sin(\pi(\nu+i\omega))}{\sin(2\pi\nu)}
\sum_{n=-\infty}^{\infty}
(-1)^n  \aMST{n}, \\
\Rmnutra&:=
 \zetainfp 2^{-s+i\omega}\omega^{\nu-s}(-i\omega)^{-\nu+i\omega}e^{-\pi \omega(1\pm 2)/2}e^{- i\pi\nu(1\pm 1)}
\frac{\sin(\pi(\nu+i\omega))}{\sin(2\pi\nu)}
\sum_{n=-\infty}^{\infty}
\frac{\Gamma(\q{n}{\nu} + \e)}{\Gamma(\q{n}{\nu}- \e)} \aMST{n},
\label{eq:Rnu-,tra} 
\end{align}
\end{subequations}
where the upper/lower signs respectively correspond to $\text{Re}(\omega)>0/\text{Re}(\omega)<0$, and where we have again used that $s\in \mathbb{Z}$. This shows again, more explicitly, that $R_+^{\nu}$ and $R_-^{\nu}$ are proportional to, respectively, 
$R^{(\infty)}_{-}$ and $R^{(\infty)}_{+}$.
Following the hatted notation in Eq.~\eqref{eq:hatted quantities},
we define $\Rpnuhat:= \Rpnu/\Rpnutra$. 

From now on and for the rest of this subsection, we choose the upper signs: i.e., we restrict ourselves to
$\text{Re}(\omega)>0$ --- one can get the results for $\text{Re}(\omega)<0$ from the symmetries in \eqref{eq:symm hat radial slns}\MC{What about $\text{Re}(\omega)=0$?}.
It is then easy to check that
\begin{equation}\label{eq:R+/- nu -> -nu-1}
R^{-\nu-1}_{+}=
\mathcal{C}_+
\Rpnu,\quad
R^{-\nu-1}_{-}= 
\mathcal{C}_-
R^{\nu}_{-},
\end{equation}
where
\begin{equation}
\mathcal{C}_+:= -(-i\omega)^{-2\nu}(i\omega)^{2\nu}e^{-2\pi i \nu},
\quad
\mathcal{C}_-:=
e^{2 \pi i \nu} \frac{\sin\left(\pi(\nu-i\omega)\right)}{\sin\left (\pi(\nu+i\omega)\right) }.
\end{equation}
The first expression can be explicitly written as
\begin{align}\label{eq:R+ nu -> -nu-1}
\mathcal{C}_+ =
\begin{cases}
  -1,& \quad 
  \arg\omega\in(0,\pi/2], \\
  -e^{-4 \pi i \nu}, &\quad  
  \arg\omega\in(-\pi/2,0]. 
  \end{cases}
\end{align}
As seen above, $R_{\pm}^{\nu}$ are solutions
of the homogeneous Teukolsky equation and so, from Eq.~\eqref{eq:R+/- nu -> -nu-1}, so are
$R_{\pm}^{-\nu-1}$, as well as $R^{-\nu-1}_{\Cinf,+}$ and $R^{-\nu-1}_{0,+}$.

From Eqs.~\eqref{eq:Rin - RC}, \eqref{eq:RC - R+/-}, and \eqref{eq:R+/- nu -> -nu-1} we have
\begin{equation}\label{eq:RinMatching}
\Rin=\left(K_{\nu}+
\mathcal{C}_+
K_{-\nu-1}\right)\Rpnu +
\left(K_{\nu}+
\mathcal{C}_-
K_{-\nu-1}
\right)R^{\nu}_{-}.
\end{equation}

Finally, using the large-$x$ asymptotics of Eqs.~\eqref{eq:Rnu+ large-x} and \eqref{eq:Rnu- large-x} in Eq.~\eqref{eq:Rin - RC} we obtain the incidence and reflection coefficients at infinity, as defined
via Eq.~\eqref{eq:Rin bc}:
\begin{equation}\label{eq:IncI coeff}
\Rininc
=
\left(K_{\nu}+
\mathcal{C}_+
K_{-\nu-1}\right)\zetainfp 2^{s-i\omega}\omega^{\nu+s}(i\omega)^{-\nu-i\omega}e^{-3\pi \omega/2}e^{-\pi i}
\frac{\sin(\pi(\nu+i\omega))}{\sin(2\pi\nu)}
\sum_{n=-\infty}^{\infty}
(-1)^n  \aMST{n},
\end{equation}
for the incidence coefficient and
\begin{align}\label{eq:RefR}
\Rinref
= & \, \left(K_{\nu}+
\mathcal{C}_-
K_{-\nu-1}
\right)
 \zetainfp 2^{-s+i\omega}\omega^{\nu-s}(-i\omega)^{-\nu+i\omega}e^{-3\pi \omega/2}e^{- 2\pi i\nu}
 \\ &\times
\frac{\sin(\pi(\nu+i\omega))}{\sin(2\pi\nu)}
\sum_{n=-\infty}^{\infty}
\frac{\Gamma(\q{n}{\nu}+\chi_s)}{\Gamma(\q{n}{\nu}-\chi_s)}
 \aMST{n}
\nonumber
\end{align}
for the reflection coefficient.

\subsubsection{Scattering amplitudes at the horizon}\label{sec:coeffs horizon}

The reader may wish to also make use of the upgoing scattering amplitudes at the horizon, $\mathcal R_{\rm up}$ and $\mathcal I_{\rm up}$.
These could be obtained from
the ingoing coefficients at infinity of the previous subsection combined with the Wronskian relations in Eq.~\eqref{eq:Whatequal,extreme} 
and, if $\omega\in\mathbb{R}$, Eq.~\eqref{eq:W[Rinhat,Ruphatcc]}.
For completeness, in this subsection we derive these  coefficients at the horizon directly---this will yield alternative expressions for these coefficients that are not readily obtainable from the coefficients at infinity combined with the Wronskian relations.

We find the upgoing coefficients at the horizon by asymptotic expansion of the matching expression~\eqref{eq:Rup from Rin}, taking $x \to 0$ while keeping $\vert k \vert $  finite.  
The small-$x$ asymptotics of $R^\nu_{0,+}$ directly follow from Eqs.~
\eqref{eq:R_0+^nu} and
the large-argument  behavior of the Kummer $M$ functions (which crucially depends on the angle of approach, which here corresponds to the argument of $k$) given in 
p.278 of~\cite{Erdelyi:1953}. The result is
\begin{align}\label{eq:R^nu_0+,x->0}
R^\nu_{0,+} \sim \zetahorp
 k^{\nu}(-ik)^{-2\nu}e^{-i \pi (2\nu+\et)/2} \Bigg(\mathcal A_\nu(-ik)^s \left( \frac{e^{i \pi \varsigma}}{-ik}\right)^{-i\omega-\nu}  e^{ik/(2x)}x^{-i\omega-2s} +  \mathcal{B}_\nu  (-ik)^{-s-i\omega+\nu} e^{-ik/(2x)} x^{i\omega} \Bigg),
\end{align}
with
$ \varsigma:= - {\rm sgn}( \text{Re} \, k)$
and
\begin{align}\label{eq:Anu}
\mathcal A_\nu &:=    \frac{\sin \left(\pi(\nu-i\omega)\right)}{\sin (2\pi \nu)} \sum_{n=-\infty}^{\infty} \frac{ \Gamma(\q{n}{\nu} + \e) }{\Gamma(\q{n}{\nu}- \e)}  \aMST{n},
\\
\mathcal B_\nu &:= - \pi \csc(2\pi \nu) \sum_{n=-\infty}^{\infty} \frac{ \Gamma(\q{n}{\nu} + \e) }{\Gamma(\q{n}{\nu}- \e) \Gamma(\q{n}{\nu}+\et)\Gamma(1-\q{n}{\nu}+\et)} \aMST{n}.
\end{align}
The complete small-$x$ asymptotics of the upgoing solution are found by  invoking the $\nu \leftrightarrow - \nu-1$ exchange symmetry of  \eqref{eq:Rup from Rin}. Making the appropriate identifications with Eq.~\eqref{eq:Rup bc} leads to the radial coefficients 
\begin{equation}\label{eq: Rup coeff}
\mathcal R_{\rm up} = \mathcal A_{\nu} ( K_{\nu})^{-1} \zetahorp e^{-\pi (2i\nu+\omega)/2} \, k^{\nu+s} (-ik)^{-2\nu} \left(\frac{e^{i \pi \varsigma}}{-ik}\right)^{-\nu-i\omega}  + (\nu \to - \nu -1)
\end{equation}
and 
\begin{equation}\label{eq: Iup coeff}
\mathcal{I}_{\rm up} = \mathcal{B}_{\nu} (K_\nu)^{-1}\, \zetahorp (-1)^se^{-\pi(2i\nu+\omega)/2} \, k^{\nu-s} (-ik)^{-\nu-i\omega} + (\nu \to -\nu-1),
\end{equation} 
where ``$(\nu \to - \nu -1)$" denotes the  term preceding it evaluated under this transformation.

\subsection{Series coefficients and renormalized angular momentum  for $\omega\to 0, m$}\label{sec:coeffs,nu omega->0,m}

Later in the paper we shall be interested in obtaining the contribution to the Green function coming from the branch points at $\omega=0$ and $\omega=m$.
For that purpose, in this subsection we analyze the series coefficients $\aMST{n}$ and the
renormalized angular momentum $\nu$ as $\omega\to 0, m$.

Let us first consider an expansion for $\eps$ small of the recurrence relations satisfied by the series coefficients $\aMST{n}$.
It is clear from Eq.~\eqref{eq:rec.Eq. coeffs} that 
$\alpha_n=O(\eps)$,
 $\beta_n=O(1)$, and
 $\gamma_n=O(\eps)$, except for possible special values of $\nu$ such that $\beta_n$ or the denominators of $\alpha_n$ or $\gamma_n$ vanish for $\eps\to 0$.
Furthermore, from Eq.~(\ref{eq:a_n large-n}), we have that $R_n=O(\eps)$
and $L_{-n}=O(\eps)$  
for sufficiently large $n$.
Therefore, except for the mentioned possible special cases, the order of $\aMST{n}$ for small $\eps$ increases as $|n|$ increases.
We will not consider these possible special cases, since in principle they should not affect the behavior of the radial solutions to {\it leading} order for small $\eps$ [see Eq.~(172)~\cite{Sasaki:2003xr}].
Therefore, barring these special cases, we have that
\begin{equation}\label{eq:an small eps}
\aMST{n}=O\left(\eps^{|n|}\right), \quad \eps\to 0.
\end{equation}

Obtaining the leading-order behavior of $\nu$ for small $\eps$ is trivial.
One just imposes that Eq.~(\ref{eq:eq for nu}) is satisfied to $O(1)$ for small $\eps$ with the choice of, e.g., $n=1$, 
and uses the property, shown above, that $R_2=O(\eps)$
and $L_{-1}=O(\eps)$.
This requires that both the $O(1)$ and $O(\eps)$ terms in $\beta_0$
are zero.
Requiring first the $O(1)$ term in $\beta_0$ to be zero readily yields
\begin{equation}\label{eq:eq nu eps=0}
\lim_{\omega\to 0, \ocr}\nu(\nu+1)=
\lim_{\omega\to 0, \ocr}\eigenbA.
\end{equation}
This result agrees with Leaver~\cite{Leaver:1986a}.
Note that if we had   used Eq.~(\ref{eq:eq for nu}) with $n=k$, for some $k\in\mathbb{Z}$, we would have found the same equation 
as Eq.~\eqref{eq:eq nu eps=0} but with $\nu$ shifted by $k$.
Therefore, fixing the value of $n$ in the  condition Eq.~\eqref{eq:eq for nu}
eliminates the symmetry of the MST formalism under $\nu\to \nu+k$, $k\in\mathbb{Z}$, 
and in this paper we choose $n=1$ in \eqref{eq:eq for nu} to determine $\nu$.

In the limit $\omega\to 0$,  we have
$\eigenbA\to \bar{A}_{\ell m 0}=\ell(\ell+1)$, and so Eq.~(\ref{eq:eq nu eps=0})
implies that $\nu=\ell$ or $\nu=-\ell-1$ for $\omega=0$.
In its turn, in the 
superradiant bound limit, Eq.~(\ref{eq:eq nu eps=0})
implies that
\begin{equation}\label{eq:nu freq critical}
\lim_{\omega\to\ocr}\nu=\nu_{c,\pm}:= -\frac{1}{2}\pm\sqrt{\frac{1}{4}+ {}_s K_{\ell m} -2m^2}, 
\end{equation}
where $\eigenK := \left(\eigenA+M^2\omega^2 + s(s+1) \right)|_{\omega=m}=\eigenAmeqo+m^2/4 + s(s+1)$ is the separation constant of \cite{Gralla:2015rpa, Casals:2016mel,Gralla:2017lto} commonly used in the Kerr/CFT literature, for example in \cite{Porfyriadis:2014fja}. 
As expected, the relationship $\nu_{c,\pm}=-\nu_{c,\mp}-1$ is satisfied.
The parameter $h$ in \cite{Casals:2016mel,Gralla:2017lto}, which determines the rate of the horizon instability of a field perturbation to extremal Kerr, is equal to
$-\nu_{c,-}$.
We also note that $-\left(\nu_{c,\pm}+1/2\right)^2$ is equal to $\delta^2$ in Eq.~(A6)~\cite{teukolsky1974perturbations} (which is the same as $\delta^2$ in the  appendix of~\cite{detweiler1980black})
for $a=M=1/2$ at $\omega=m$.
As mentioned in Sec.\ref{sec:coeffs},
it has been observed that, for $\omega$ real, $\nu$ 
is either real or else it is complex with a real part equal to a half-integer number. Similar properties were observed by us in Eq.~(23) of~\cite{Casals:2016mel} and Eq.~(67) of~\cite{Gralla:2017lto} for  $h$.
Finally, a property which we shall utilize  later is that $\nu_{c,\pm}$ 
is invariant under $s\to -s$, 
since $\eigenK=\eigenKms$  follows from Eq.~\eqref{eq:symms eigen}.

Requiring  the $O(\eps)$ term in $\beta_0$ to be zero would yield
the next-to-leading order term for $\nu$.
Orders higher than leading order are easy to find, then, by introducing  in 
Eq.~(\ref{eq:eq for nu}) an  expansion
for $\bar{A}_{\ell m}$ for either $\omega$ small or $k$ small, which is assumed to be known (this is certainly true for $\omega$ small---see~\cite{seidel1989comment}) and a corresponding expansion
for $\nu$ with undetermined coefficients.
These coefficients can be found by iteratively solving the equation to higher orders. 
For example, for small $|\omega|$ (and any $\arg\omega$), we have
\begin{align}\label{eq:nu omega->0}
&
\nu=\ell+\nu_2\omega^2+ \nu_3\omega^3
+ O\left(\omega ^4\right),
\end{align}
where
\begin{subequations}\label{eq:nu_2,3}
\begin{align}
\nu_2:=\ &\frac{\left(-15 \ell^4-30 \ell^3-6 \ell^2 s^2-4 \ell^2-6 \ell s^2+11 \ell-3 s^4+6 s^2\right)}{2 \ell (\ell+1) (2 \ell+1) \left(4 \ell^2+4
   \ell-3\right)},
 \\
  \nu_3:= \ & 
   \frac{m}{(\ell-1) \ell^2 (\ell+1)^2 (\ell+2) (2 \ell-1) (2 \ell+1) (2 \ell+3)}
   \Big(5 \ell^6+15 \ell^5+\ell^4 \left(3 s^2+2\right)+3 \ell^3 \left(2 s^2-7\right) +
 \nonumber \\ &
   \ell^2 \left(3 s^4-6 s^2-7\right)+3 \ell \left(s^4-3 s^2+2\right)+s^2
   \left(5 s^4-16 s^2+11\right)\Big).
\end{align}
\end{subequations}

\section{Analytical properties of the transfer function}\label{sec:analytical properties}

In pioneering work~\cite{Leaver:1986}, Leaver
deformed the Laplace integral contour corresponding to Eq.~\eqref{Eq:TDGF}  in Schwarzschild spacetime into the complex frequency plane.
In doing so, he revealed how various types of singularities in the transfer function contribute to the full Green function in their own ways.  Subtleties aside, the qualitative picture drawn by Leaver goes as follows. At ``early'' times, direct propagation on the future light cone derives from a large-$|\omega|$ arc (the only contribution that does not come from a singularity in the transfer function).  At very late times, the field exhibits a power-law time dependence deriving from frequencies near the branch point(s) on the real axis. At ``intermediate'' times, the field takes the form of a decaying sinusoid coming from the  quasinormal modes\footnote{Although perhaps less known, there is also a contribution at intermediate times from frequencies along the branch cut which are not ``near" the branch point~\cite{CDOW13,Casals:2012ng}}.

Indeed, Leaver's picture is supported by the asymptotic theory of Laplace transforms  \cite{Doetsch1974}, in which the late-time behavior of a function  of time, say $f(t)$ as $t\to\infty$, is related to the asymptotics of its Laplace transform, $\tilde f(\omega)$, near its uppermost singular point $\omega_0$, or points $\{\omega_i\}_{i=0,1,2,\ldots}$ if more than one happen to lie on the same abscissa, in the complex-$\omega$ plane. 

As mentioned in Sec.\ref{sec:radial series}, in extremal Kerr, the radial solutions $R^{(\infty)}_{\pm}$ in principle possess a branch point at  zero frequency ($\omega=0$) and  $R^{(0)}_{\pm}$ possess a branch point at the superradiant bound frequency ($\omega=m$).
As we shall explicitly show in the following sections, 
the radial solutions indeed possess
these branch points and they carry over to the transfer function. Whereas the branch point at $\omega=0$ already exists in subextremal Kerr,
with associated late-time decay having been analyzed in \cite{Leaver:1986,PhysRevLett.84.10,casals2016high}, the branch point at $\omega=m$ is new to extremal Kerr. The late-time decay of the master field on the horizon due to the emergent branch point at $\omega=m$ has been analyzed in Refs.~\cite{Casals:2016mel,Gralla:2017lto}.

Apart from the ``physical" branch points at $\omega=0$ and $m$ that give rise to the late-time decay in Leaver's picture, the various mode quantities may also possess other branch points which may be deemed ``unphysical" in a certain sense.
These unphysical branch points are of two types.
The first type 
of unphysical branch points comes from the angular eigenvalue $\eigenA$ and eigenfunction ${}_sS_{\indmode}$ \cite{oguchi1970eigenvalues,hunter1982eigenvalues,BONGK:2004}, and in principle carry over to 
the renormalized angular momentum $\nu$ and the MST series coefficients $\aMST{n}$.
It has been shown that the angular branch points do not lie on the real axis when $s=0$~\cite{hunter1982eigenvalues}. Numerical evidence suggests that this is also the case for $s\neq 0$ (for this, one can make use of the \rm{MATHEMATICA} toolkit in~\cite{BHPToolkit}). Furthermore, it has been shown that the angular branch points vanish upon summation over $\ell$ and $m$~\cite{casals2016high}, contributing nothing to the full Green function. 
The second type of unphysical branch points is observed directly
in $\nu$ and $\aMST{n}$.
Numerical evidence presented in~\cite{MSc-Longo,Casals:Longo:2018} suggests that these two quantities possess discontinuities which
may be removed by using 
the above ``MST symmetries" of addition of an integer to $\nu$ and/or  transformation $\nu\to -\nu-1$.
Other than singularities of the removable type and possible angular  branch points (inherited from the eigenvalue), $\nu$ and $\aMST{n}$ do not appear to display any further discontinuities.
In conclusion, we are confident that potential discontinuities in
$\eigenA$, ${}_sS_{\indmode}$, $\nu$ and $\aMST{n}$
will not influence any physical quantities such as the scalar/electromagnetic/gravitational wave tail. They will thus be ignored in subsequent calculations. 

\begin{figure}
  \centering
  \includegraphics[width=.6\textwidth]{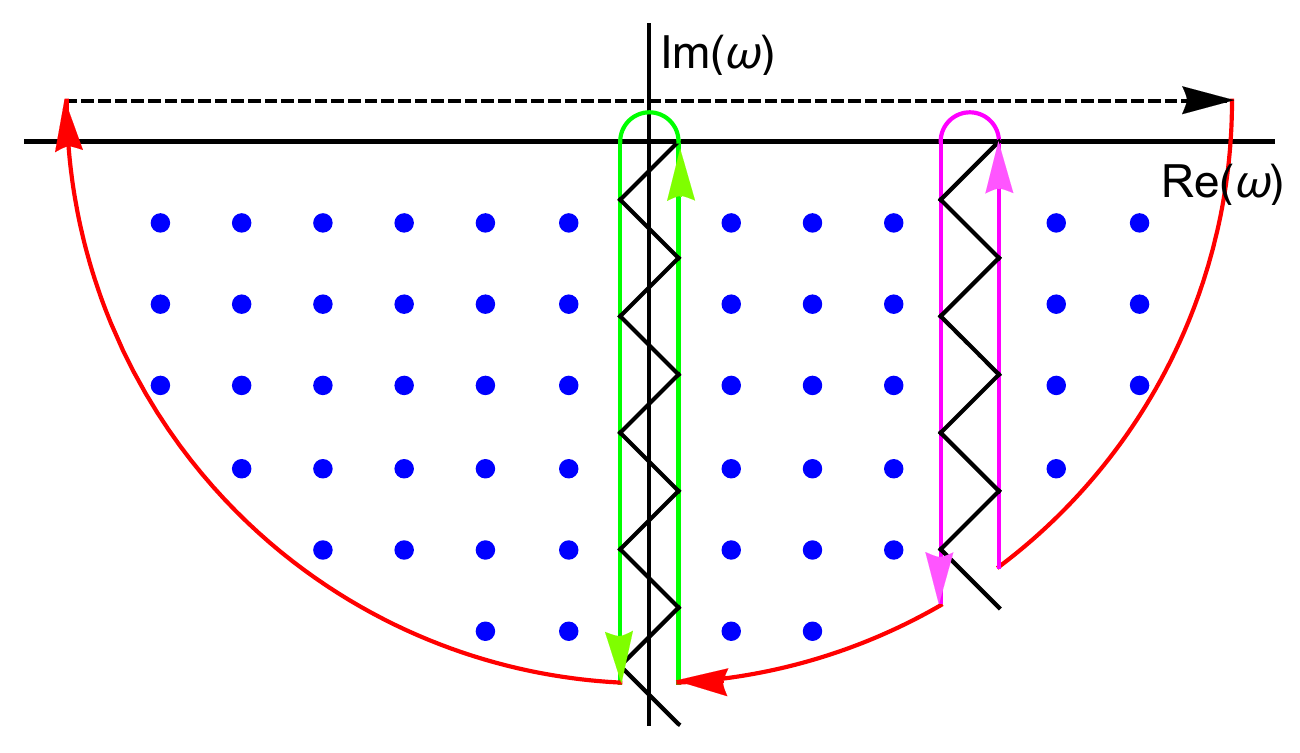}
  \caption{Schematic representation in the complex frequency plane of the singularities of the transfer function and integration contours for extremal Kerr.
  The physical singularities are as follows: (i)  blue dots correspond to simple poles (quasinormal modes; for simplicity, we plot them symmetrically with respect to the negative imaginary axis, although that is generally not the case in Kerr spacetime); (ii) two crisscrossed black lines corresponding to branch cuts down from the origin $\omega=0$ and the superradiant-bound frequency $\omega=m$ (here represented for $m>0$). 
  The dashed (black) straight line corresponds to the original integration contour in Eq.~\eqref{Eq:TDGF}.
  This contour may be deformed so as to yield an integration over a high-frequency arc (red semicircle) together with integrals around 
the two branch cuts (green and pink contours wrapped around the crisscrossed lines).
N.B.: here we omit any ``unphysical'' (see text) branch points in the transfer function.
}
  \label{fig:compl freq}
\end{figure}

Apart from the above mentioned physical branch points at $\omega=0$ and $m$, the transfer function also possesses poles in the complex-frequency plane corresponding to the quasinormal modes
In Fig.\ref{fig:compl freq} we schematically represent the various physical singularities  of the transfer function in the complex-frequency plane, as well as the integration contours for the Green function.

There is strong numerical evidence ~\cite{Richartz:2016,PhysRevD.97.061502,Casals:Longo:2018} that the uppermost singular points of the transfer function in extremal Kerr are the  branch points
at  the origin  ($\omega=0$) and the superradiant bound ($k=0$). A complementary analytical argument may be found in Sec.VII.B of~\cite{Casals:Longo:2018}.
Furthermore, there is a rigorous result for mode stability in the very recent~\cite{costa2019mode}
and rigorous linear stability results exist for the specific case of {\it axisymmetric} scalar field perturbations in~\cite{aretakis2012decay}.

Using our MST expressions, in the next three sections we provide a formalism for obtaining the discontinuity in the transfer function across the branch cuts (BCs) extending from $\omega=0$ and from $k=0$ to arbitrary order in the frequency. We explicitly calculate the leading-order tail due to the 
$\omega=0$ branch point and the  leading-order transfer function near $k=0$ (which yields the known leading order tail due to the $k=0$ BC).
The former is a new result, whereas we find that the latter agrees with the existing results in~\cite{Casals:2016mel,Gralla:2017lto} obtained using MAE.

\section{Branch cut from the origin and tail }\label{sec:BC origin}

As advanced in Sec.\ref{sec:radial series}, in extremal Kerr, the upgoing radial solution possesses a branch point at $\omega=0$ in the complex frequency plane, which carries over to the transfer function, similarly to what happens in subextremal Kerr.
We choose the corresponding BC to point down the negative imaginary axis.

We shall use the notation that, if $A=A(\omega)$ is a function of the frequency possessing a branch point at $\omega=0$, then
\begin{equation}
\Disc A := A(\omega)-A\left(\omega e^{2\pi i}\right)
\qquad \text{(change in function } A \text{ across BC extending from } \omega=0).    
\end{equation}

In particular, the discontinuity in the transfer function across the BC
is
\begin{align}
\Disc\tilde{g}_{\ell m\omega}(x,x'):= 
\left.\tilde{g}_{\ell m \omega}(x,x')\right|_{\omega=-i\fNIA}
-\left.\tilde{g}_{\ell m\omega}(x,x')\right|_{\omega=-i\fNIA e^{2\pi i}},
\end{align}
where $\fNIA>0$.
The 
contribution to
an $(\ell,m)$-mode of the Green function due to the branch point at the origin is then obtained by, essentially, integrating the discontinuity in the transfer function, $\Disc\tilde{g}_{\ell m\omega}(x,x')$, along the corresponding BC:
\begin{align}\label{eq:delta G}
\Disc \Glm(x^{\mu},x^{\mu'}):= -i\, e^{i m\phi}
\int_{0}^{\infty}d\fNIA\
\left.e^{-\fNIA t}
\Disc\tilde{g}_{\ell m\omega}(x,x')
 {}_s\eigenSS_{\indmode}(\theta,\theta') \right|_{\omega=-i\fNIA}.
\end{align}
As explained earlier, the late-time behavior of $\Disc \Glm$ will be given by the small-$\fNIA$ behavior of the integrand in Eq.~\eqref{eq:delta G}.

In this section, we  lay out the MST formalism for calculating the discontinuity of the transfer function  across  the 
BC down from the origin.
We also calculate,
in separate subsections,
the contribution of this BC to leading order for late times for a field point at: (i) timelike infinity ($t\to \infty$, $r_*$ finite), (ii) the future event horizon ($u\to \infty$, $v$ finite) and (iii) future null infinity ($v\to\infty$, $u$ finite),
 where 
$u:= t-r_*$
is retarded time,
$v:= t+r_*$
is advanced time and
$r_* := x-1/(2x)+\ln x$.

\subsection{Discontinuity in the up modes} \label{sec:disc up}
Here, we give an analytic expression for the discontinuity in the up solution across the BC originating from $\omega=0$.
For this purpose, we will adapt to extremal Kerr a technique used in~\cite{Casals:Ottewill:2015} in Schwarzschild spacetime and in~\cite{casals2016high} in subextremal Kerr spacetime.

Our starting point is Eq.~(\ref{eq:Rup}) for the up radial solution.
In it, we identify the source of the discontinuity across the BC down the negative imaginary axis. There are two factors 
(leaving aside $\zetainfp$)
that are discontinuous across this cut: $\omega^\nu$ and the confluent hypergeometric function $U$.
The analytic continuation of the first factor is trivial:
\begin{equation}
\omega^{\nu} \to e^{2\pi i\nu}\omega^{\nu}\quad\text{as}\quad\omega\to e^{2\pi i}\omega.
\end{equation}
The analytic continuation of the second factor is given in Eq.~(13.2.12)~\cite{NIST:DLMF}.
Combining the analytic continuation of the two factors, we can write
{\small
\begin{align}\label{eq:Rup anal cont}
\frac{\Rup}{\zetainfp} \Bigg\vert_{\omega\to e^{2 \pi i} \omega}
=
\frac{f_{\rm up}(x,\omega)}{\zetainfp(\omega)}
e^{2\pi i\nu}
\sum_{n=-\infty}^{\infty}
A^{up}_n(x,\omega)
\left(\left(1-e^{-2\pi i b}\right)\frac{\Gamma(1-b)}{\Gamma(1+\aaa-b)}M(\aaa,b,-2i\omega x)+
e^{-2\pi i b}U(\aaa,b, -2 i \omega x)
\right),
\end{align}}
where
\begin{equation}\label{eq:a,b}
\aaa:= \q{n}{\nu} + \e, \quad b :=  2\q{n}{\nu}.
\end{equation}
We find it convenient to normalize the up solution by its transmission coefficient, i.e., to use $\Ruphat$ instead of $\Rup$. The analytic continuation of this coefficient 
(``normalized" via $\zetainfp$) follows trivially from
\eqref{eq:Ruptra}:
\begin{equation}\label{eq:Ruptra cont}
\left.\frac{\Ruptra}{\zetainfp}
\right|_{\omega\to e^{2 \pi i} \omega}
=
e^{-2\pi\omega}
\frac{\Ruptra}{\zetainfp}.
\end{equation}
We next use Eq.~(6.7.7) of Vol.~1~\cite{Erdelyi:1953}
to express the $M$ function in Eq.~\eqref{eq:Rup anal cont}
in terms of $U$ functions,
as well as the following straightforward identity:
\begin{equation}\label{eq:simpl}
e^{2\pi i \nu}\left(1-e^{-2\pi i b}\right)
\frac{\Gamma(1-b)\Gamma(b)e^{-\pi i a}}{\Gamma(1+\aaa-b)\Gamma(b-\aaa)}=
e^{-2\pi\omega}-e^{-2\pi i\nu},
\end{equation}
in order to obtain
\begin{align}\label{eq:Rup/T anal cont}
\Disc 
\Ruphat
=\frac{f_{\rm up}}{\Ruptra}
e^{-2i\omega x}e^{2\pi (\omega+i\nu)}
\left(e^{-2\pi i\nu}-e^{-2\pi\omega}\right)
\sum_{n=-\infty}^{\infty}
(-2i\omega x)^n\aMST{n}U(b-\aaa,b,2i\omega x).
\end{align}
With all the quantities on the right-hand side 
of Eq.~\eqref{eq:Rup/T anal cont}
evaluated just on the right side of the BC, i.e., for $\omega=\lim_{c\to 0^+}( -i\fNIA +c)$,
assuming $\fNIA>0$ throughout,
we have  an analytic expression for the discontinuity across the negative imaginary axis of the up solution normalized to have unit transmission coefficient.
By comparison with Eq.~(\ref{eq:R_+^nu}), we observe that this discontinuity
is proportional to the solution $\Rpnu$,
\begin{equation}\label{eq:discont Rp/T-R+}
\Disc
\Ruphat = i\, q(\fNIA)\,
\left.\Rpnuhat\right|_{\omega=-i\fNIA},
\end{equation}
where the BC ``strength" 
$q(\fNIA)$ is given  by
\begin{equation}\label{eq:q}
q(\fNIA):= -2\sin\left(2\pi\nu\right)e^{2\pi \omega}\frac{\Rpnutra}{\Ruptra},
\end{equation}
with all quantities on the right-hand side evaluated at $\omega=\lim_{c\to 0^+}( -i\fNIA +c)$, with $c>0$.
We note that $\Rpnuhat$ has a branch point at $\omega=0$. But, because of the boundary conditions Eq.~\eqref{eq:Rnu+ large-x}  that it satisfies at infinity
(with a wave
term $e^{-i\omega(x+\ln x)}$ being the complex-conjugate of the wave term
$e^{+i\omega(x+\ln x)}$
in the boundary conditions for $\Ruphat$), 
its BC lies on the {\it positive} imaginary axis. Since $\Rpnuhat$ in Eq.~\eqref{eq:discont Rp/T-R+}  is evaluated on the {\it negative}
imaginary axis, there is no ambiguity as to its value there.

The proportionality relation in Eq.~\eqref{eq:discont Rp/T-R+} 
was to be expected for the following reason.
The function $\Ruphat$ is a  solution of the homogeneous
radial equation having the same (purely outgoing, $e^{+i\omega(x+\ln x)}$) asymptotic behavior at radial infinity whether it is evaluated at $\omega$ or at $\omega e^{2\pi i}$.
As mentioned in Sec.\ref{sec:overview}, however, this behavior corresponds to an exponentially dominant solution when $\text{Im}(\omega)<0$, as is the case for $\omega$ on the negative imaginary axis.
While the solutions $\Ruphat$  at $\omega$ and at $\omega e^{2\pi i}$ have the same dominant asymptotic behavior, they differ in the amount of the subdominant solution (which is purely ingoing, $e^{-i\omega(x+\ln x)}$, at radial infinity) that they contain.
This means that  $\Disc \Ruphat$
is a solution of the homogeneous radial equation 
and 
satisfies a purely-ingoing
boundary condition  at radial infinity (and for $\omega$ on the negative imaginary axis).
Since $\Rpnu$ is also a solution of the homogeneous radial equation and, from \eqref{eq:Rnu+ large-x}, it is purely ingoing at radial infinity, it follows that it 
must be proportional to $\Disc \Ruphat$.
Equation \eqref{eq:discont Rp/T-R+} is this proportionality relation.
This expression for the discontinuity of the upgoing radial solution across the BC allows us to find the corresponding discontinuity of the transfer function that we give in the next subsection.

\subsection{Discontinuity in the transfer function} \label{sec:disc GF omega=0}

In~\cite{casals2016high} it was shown that the discontinuity in the transfer function across the BC is given by
\begin{align} \label{Eq:GBC}
\Disc \glmo(x,x')=-2 i \fNIA \frac{q(\fNIA)}{\Wbp\Wbm}\Big[ \Rinhat(x) \Rinhat(x') \Big]_{\omega=-i\fNIA}.
\end{align}
Here, $\Whatpm$ is equal to $\What$ evaluated, respectively, on the right/left of the BC and, as always, $\fNIA>0$.
Equation \eqref{Eq:GBC} was proven 
in~\cite{casals2016high} in subextremal Kerr and with the
subextremal counterparts of the Wronskian $\What$  and the BC strength $q$. However, it is trivial to see that Eq.~\eqref{Eq:GBC} is also valid in extremal Kerr with the Wronskian $\What$  and BC strength $q$ as defined in Eq.~\eqref{eq:discont Rp/T-R+}, which is used in order to derive Eq.~\eqref{Eq:GBC}.

We note the following pertinent point for the small-$\omega$ asymptotics of the transfer function. 
Infinite series of confluent hypergeometric $U$ functions whose last argument goes to zero as $\omega\to 0$ and whose second argument grows with $n$ at least like $2n$, 
such as the series for $\Rup$ in Eq.~\eqref{eq:Rup}
and for $\Rpnu$ in Eq.~\eqref{eq:R_+^nu}, are not amenable to asymptotics for $\omega\to 0$ (see, e.g., Eq.~(13.2.16)~\cite{NIST:DLMF} together with Eq.~\eqref{eq:an small eps}).
On the other hand, Eq.~\eqref{Eq:GBC} offers an
expression for $\Disc \glmo$  as proportional to the ingoing solution: $\Rin(x)\Rin(x')$. The ingoing solution has the representation Eq.~\eqref{eq:Rin} containing
$U$ functions whose last argument does not go to zero as $\omega\to 0$ and is thus amenable to $\omega\to 0$ asymptotics.
This allows us to find the small-$\omega$ asymptotics of $\Disc \glmo$.

An expression for the denominator in Eq.~(\ref{Eq:GBC}) in terms of the ingoing radial coefficients is also given in~\cite{casals2016high}\footnote{We note that there is a typographical error in Eq.~(5.10)~\cite{casals2016high}: the sign of the second term should be $+$ instead of $-$.}(it is valid in extremal Kerr as 
well as subextremal Kerr):
\begin{align}
\Wbp \Wbm=
\left(2 \fNIA \frac{\Rininc}{\Rintra}\right)^2+4 i \fNIA^2 q(\fNIA) \frac{\Rininc \Rinref}{(\Rintra )^2},
\label{Eq:WpWm}
\end{align}
where all the quantities are meant to be evaluated in the limit to the negative imaginary axis from the fourth quadrant.

Here we have laid the foundation for obtaining the discontinuity of the transfer function $\tilde{g}_{\ell m\omega}$ across the negative imaginary axis for any $\fNIA:=i\omega>0$.
From  Eqs.~\eqref{Eq:GBC}, \eqref{Eq:WpWm}, \eqref{eq:q}, and the expressions for the radial coefficients and for $\Rinhat$ derived  in the previous sections, one can obtain the discontinuity
of the transfer function across the BC down from $\omega=0$ and so, via Eq.~\eqref{eq:delta G}, the  late-time tail of the $(\ell,m)$-modes of the retarded Green function.
This could be done either exactly, e.g., via a semianalytic/numeric evaluation 
of the infinite sums  or analytically up to arbitrary order at late times
by systematically expanding the expressions for small
frequency, in a manner similar to~\cite{casals2016high} in subextremal Kerr or to~\cite{Casals:Ottewill:2015} in Schwarzschild\footnote{See also~\cite{Casals:2012ng,Casals:2011aa}, where use is made of other high-order techniques, which could be adapted to Kerr.}. 
In the next subsections we shall do the latter to leading order: we provide the leading small-$\omega$ behavior of various quantities and use them to obtain the
leading late-time behavior of an $(\ell,m)$-mode of the 
Green function.
We note that the results in these subsections are not valid in the axisymmetric case ($m=0$), where
the branch points at the origin and at the critical frequency
coincide at $\omega=0$.
Reference ~\cite{Gralla:2017lto} gives the late-time asymptotics on the horizon in this axisymmetric case.

\subsection{Small-$\omega$ asymptotics of radius independent quantities appearing in $\delta \glmo$}\label{sec:small omega nec for dg}
In order to compute the late-time behavior of the contribution to the Green function due to the branch point at $\omega=0$, we require the small-$\omega$ asymptotics of the radius-independent constituents of $\delta \glmo$ in Eq.~\eqref{Eq:GBC}, namely, $q(\sigma)$ and $\mathcal W^+ \mathcal W^-$. Here we provide the necessary asymptotic expressions for these terms.     

First of all, from Eq.~\eqref{eq:Knu} we obtain the needed asymptotics for the matching coefficients 
\begin{align}\label{eq:Knu,omega->0}
K_{\nu}\sim
\frac{\zetahorp}{\zetainfp}
(-1)^{\ell+1}m^{-\ell}2^{-\ell}\nu_2\frac{\Gamma(2\ell+1)\Gamma(2\ell+2)}{\Gamma^2(\ell+1-s)}
\omega^{-\ell},
\quad
K_{-\nu-1}
\ll K_{\nu},\quad \omega \to 0,
\end{align}
where we have used
\begin{align}
\Gamma\left(1-\q{0}{\nu}+\e\right) \sim\frac{(-1)^{\ell+s}i}{\Gamma\left(1+\ell-s\right)\omega},
\quad
\Gamma\left(1-2\q{0}{\nu}\right)\sim
\frac{1}{2\Gamma\left(2\ell+2\right)\nu_2\omega^2},
\quad \omega \to 0,
\end{align}
and $\nu_2$ is given in Eq.~\eqref{eq:nu_2,3}.

With $K_\nu$ as $\omega\to 0$ at hand, the small-$\omega$ asymptotics of the scattering amplitudes follow from
Eqs.~\eqref{eq:Ruptra}, \eqref{eq:IncI coeff}, \eqref{eq:RefR}, \eqref{eq:Rnu+,tra} and \eqref{eq:Rintra}. From these equations we obtain, respectively,
\begin{equation}\label{eq:Ruptra small-freq}
\Ruptra\sim
\zetainfp
e^{-i\pi \ell/2}
2^{-s}
\frac{\Gamma(\ell+1+s)}{\Gamma(\ell+1-s)}
\omega^{-s+i\omega},\quad \omega \to 0,
\end{equation}

\begin{equation}\label{eq:Rininc small-freq}
\Rininc\sim 
\zetahorp
m^{-\ell}2^{-\ell+s-1}e^{i\pi (1-\ell)/2}\frac{\Gamma(2\ell+1)\Gamma(2\ell+2)}{\Gamma^2(\ell+1-s)}
\omega^{-\ell-1+s-i\omega},
\quad \omega \to 0,
\end{equation}
\begin{equation}
\Rinref\sim
-\zetahorp e^{i\pi (1+\ell)/2} m^{-\ell}2^{-\ell-s-1}\frac{\Gamma(2\ell+1)\Gamma(2\ell+2)\Gamma(\ell+1+s)}{ \Gamma^3(\ell+1-s)}
\omega^{-\ell-s-1+i\omega}
,\quad \omega \to 0,
\end{equation}
\begin{equation}
\Rpnutra\sim
\zetainfp e^{i\pi (\ell-1)/2}  \frac{2^{s-1}}{\nu_2}\omega^{s-i-1\omega}
,\quad \omega \to 0,
\end{equation}
and
\begin{equation}\label{eq:Rintra,omega->0}
\Rintra
 \sim \zetahorp 
(-1)^s e^{-i\pi \ell/2}
m^s \frac{\Gamma(\ell+1+s)}{\Gamma(\ell+1-s)} 
 ,\quad \omega \to 0,
\end{equation}
where we have used
\begin{align}\label{eq:sine ratio}
\frac{\sin\left(\pi\left(\nu+i\omega\right)\right)}{\sin\left(2\pi\nu\right)}\sim \frac{(-1)^{\ell}i}{2\nu_2 \omega},\quad \omega \to 0,
\end{align}
as well as Eq.~\eqref{eq:Knu,omega->0} for $\Rininc$ and $\Rinref$.

Equation \eqref{eq:Wequal} gives the scaled Wronskian in terms of the up transmission coefficient
and the in incidence coefficient, defined via Eqs.~\eqref{eq:Rup bc} and \eqref{eq:Rin bc}, with small-$\omega$ asymptotics
given in Eqs.~\eqref{eq:Ruptra small-freq} and \eqref{eq:Rininc small-freq}.
It follows that the asymptotics for the scaled Wronskian are \footnote{Although the Wronskian  possesses a BC down from $\omega=0$, its discontinuity only appears at higher order in $\omega$, not
at the leading order of Eq.~\eqref{eq:W w->0} -- for the explicit details (such as the order in which it appears) in the case of subextremal Kerr, see~\cite{casals2016high}.}
\begin{align}\label{eq:W w->0}
\W\sim 
(-1)^{\ell+1}\zetainfp \zetahorp \left(2m\right)^{-\ell}\frac{\Gamma(2\ell+1)\Gamma(2\ell+2)\Gamma(\ell+1+s)}{\Gamma^3(\ell+1-s)}
\omega^{-\ell},
\quad \omega \to 0.
\end{align}

Lastly, from the various asymptotics above  we obtain the asymptotics for the BC strength $q$ in Eq.~\eqref{eq:q} and the Wronskian factor $\mathcal W^+ \mathcal W^-$ in Eq.~(\ref{Eq:WpWm}):
\begin{equation}\label{eq:q small-omega}
q(\fNIA)\sim 
e^{-\pi i(\ell-1/2)}2^{1+2s}\pi \frac{\Gamma(\ell+1-s)}{\Gamma(\ell+1+s)} \omega^{2s-2i\omega+1}
,\quad \omega \to 0,
\end{equation}
and
\begin{equation}\label{Eq:WpWm small-omega}
\Wbp \Wbm\sim
 (-1)^{s}2^{2(s-\ell)}m^{-2(\ell+s)} \frac{\Gamma^2(2\ell+1)\Gamma^2(2\ell+2)}{\Gamma^2(\ell+1-s)\Gamma^2(\ell+1+s)}\omega^{2(s-\ell-i\omega)}
,\quad \omega \to 0.
\end{equation}

\subsection{Late-time $\omega=0$ tail at finite  radii}\label{sec:omega->0 tail,finite}

The last quantity in Eq.~\eqref{Eq:GBC} for which we need the asymptotics is the ingoing radial function.
Given the small-$\omega$ behavior of the $\aMST{n}$ in Eq.~\eqref{eq:an small eps},
it follows that the leading-order coefficient in the $n$-sum
in
Eq.~(\ref{eq:Rin})
is given by the $n=0$ term as long as $x$ is finite.
Thus, from Eqs.~(\ref{eq:Rin}), \eqref{eq:an small eps} and (\ref{eq:nu omega->0}), we straightforwardly obtain, for $x$ finite,
\begin{equation}\label{eq:Rin omega->0}
\Rin \sim
\zeta^{(0)}_+ e^{ i \pi (s+1)/2}
x^{-s-\ell-1} 
e^{-\frac{im}{2x} } m^{\ell+1}
\frac{\Gamma(\ell+s+1)}{\Gamma(\ell-s+1)} U\left(\ell+1-s, 2\ell +2, \frac{im}{x}\right) , \quad \omega \to 0.
\end{equation}

We are now poised to carry out the integral along the BC of the small-frequency asymptotics of the transfer function. 
From Eq.~(\ref{Eq:GBC}),  together with the asymptotics
in Eqs.~\eqref{eq:Rintra,omega->0},\eqref{eq:q small-omega}, \eqref{Eq:WpWm small-omega} and \eqref{eq:Rin omega->0},
we find
\begin{align}\label{eq:delta tilde g,omega->0}
&
\Disc\tilde{g}_{\ell m\omega}\sim i\, g^{(f)}_{\ell m}(x,x')\, \fNIA^{2\ell+2},\quad \omega\to 0,
\end{align}
where
\begin{align}\label{eq:gf}
g^{(f)}_{\ell m}(x,x'):= &
(-1)^{\ell+s}e^{\pi i (s+1)/2}
\pi 2^{2(\ell+1)}m^{2(2\ell+1)}\frac{\Gamma^3(\ell-s+1)\Gamma(\ell+s+1)}{\Gamma^2(2\ell+1)\Gamma^2(2\ell+2)}(x\cdot x')^{-\ell-s-1}
e^{-im(1/x+1/x')/2}
\nn \\&
\times U\left(\ell+1-s,2\ell+2,\frac{im}{x}\right)U\left(\ell+1-s,2\ell+2,\frac{im}{x'}\right)
\end{align}
is independent of $\sigma$ and symmetric under interchange of $x$ and $x'$.

Finally, performing the integration in Eq.~\eqref{eq:delta G} using \eqref{eq:delta tilde g,omega->0}, we find
\begin{align}
\Disc \Glm
\sim 
\Gamma(2\ell+3)
e^{i m\phi}
{}_s\eigenSS_{\ell m0}
(\theta,\theta')
g^{(f)}_{\ell m}(x,x')t^{-3-2\ell},
\qquad t \to \infty, \quad(x \text{ and } x' \text{ finite})
\label{eq:delta Glm,t inf x finite}
\end{align}
for the leading late-time behavior of the  Green function $(\ell,m)$-modes due to the branch point at $\omega=0$.
The angular factor evaluated at the origin, 
${}_s\eigenSS_{\ell m0}$, so
reduces to a product of the 
well-known spin-weighted {\it spherical} harmonics~\cite{goldberg1967spin,newman1966note}.
We note that the result in Eq.~\eqref{eq:delta Glm,t inf x finite} is valid for a field point approaching timelike infinity and a source point at an arbitrary finite radius away from the horizon (since we
have kept $r$ and $r'$ fixed and finite while taking $t\to\infty$).

\subsection{Late-time $\omega=0$ tail at radial infinity}\label{sec:omega->0 tail,inf}

The results of the previous subsection are not valid as the field point $x$ approaches radial infinity. 
In this section we modify the asymptotic analysis in order to obtain the late-time behavior when $x\to \infty$ and $x'$ is finite. As opposed to the finite radii case above, here we do not use directly the formalism of Sec.\ref{sec:disc GF omega=0}. Instead, we use
the original expression for the transfer function, i.e., Eq.~(\ref{eq:radialgreenfunction}), and find the asymptotics as $\omega\to 0$, while $x \omega \to \infty$ for each quantity in the expression. This can be accomplished 
by taking a parameter $\lambda$ to zero while fixing $\bar x := \lambda^{3/2} x$ and $\bar \omega := \omega/\lambda$, say.

For the in solution, Eq.~\eqref{eq:Rin omega->0} shows that it has a finite limit as $\omega \to 0$ and contributes only to the overall amplitude of the late-time behavior due to the corresponding BC.
That is, at arbitrary finite radius we have 
\begin{equation}
\Rin =  O(1), \quad \omega \to 0.
\end{equation}

Given the small-$\omega$ behavior of the $\aMST{n}$ in Eq.~\eqref{eq:an small eps}, of $\nu$ in Eq.~\eqref{eq:nu omega->0}, and the small-$\omega$ but large-$x \omega$ behavior of the confluent hypergeometric function $U$ in Eq.~(13.7.3)~\cite{NIST:DLMF},  it follows that the leading-order term in the $n$-sum in Eq.~(\ref{eq:Rup}) for $\Rup$ is given by the $n=0$ term.
From Eqs.~(\ref{eq:Rup}), \eqref{eq:an small eps} and \eqref{eq:nu omega->0}, we straightforwardly obtain, as $\omega \to 0$ with $x$ large and $x\omega$ not necessarily small,
\begin{equation}
\Rup\sim
R^{\rm up}_0
 \omega^{\ell+1} U(\ell+1 - i \omega +s, 2 \ell +2, -2 i \omega x),
\label{eq:Rup omega->0}
\end{equation}
where
\begin{equation}
R^{\rm up}_0:=
\zeta_+^{(\infty)}2^{\ell+1}
\frac{\Gamma\left(\ell+s+1\right)}{\Gamma\left(\ell-s+1\right)}
x^{-s+\ell} e^{-i \pi  s/2} e^{ - \pi i (\ell+1/2)}  e^{i\omega x}. 
\end{equation}
We note that although we could have applied Eq.~(13.7.3)~\cite{NIST:DLMF} already at this stage, we find it 
better to compute the discontinuity in $\Rup$ before applying the large-$\omega x$ asymptotics.

The discontinuity across the BC down from $\omega=0$, to leading order for small frequency, is then
due to the small-frequency asymptotics of the up solution, Eq.~\eqref{eq:Rup omega->0}.
From Eq.~(\ref{eq:Rup omega->0}) and the analytic continuation of Eq.~(13.2.12)~\cite{NIST:DLMF}, together with Eq.~(13.7.2)~\cite{NIST:DLMF}, it follows that
\begin{align} 
\delta\Rup &\sim  R^{\rm up}_0 \omega^{\ell+1}\Big(
U(\ell+1 - i \omega +s, 2 \ell +2, -2 i \omega x)-U\left(\ell+1 - i \omega +s, 2 \ell +2, -2 i \omega e^{2\pi i} x \right)\Big), \,\,\omega\to 0 \nonumber, \\ 
&\sim  R^{\rm up}_0 \omega^{\ell+1}\, \frac{2\pi(-1)^{s-\ell+1}\Gamma(\ell-s+1)}{\Gamma(2\ell+2)}M(\ell+1+s,2\ell+2,-2i\omega x), \,\,\omega\to 0, \\
&\sim 2\pi R^{\rm up}_0(-2ix)^{-\ell-1-s}\omega^{1-s+i\omega}, \qquad\,\,\omega\to 0, \,\,\, x\omega\to\infty,
\label{eq:discont Rup lead}
\end{align}
where we have discarded the subdominant term in the asymptotics proportional to $e^{-2i\omega x}$, which is exponentially suppressed when evaluated on the negative imaginary axis where we compute the BC integral.

We can now carry out the integral along the BC of the small-frequency asymptotics for the transfer function. 
From Eq.~(\ref{eq:radialgreenfunction}), and putting together the asymptotics
of Eqs.~\eqref{eq:discont Rup lead}, \eqref{eq:W w->0}, and \eqref{eq:Rin omega->0},
we obtain
\begin{align}\label{eq:delta tilde g,omega->0,x omega->inf}
\Disc\tilde{g}_{\ell m\omega}(x,x')\sim i\,g^{(\infty)}_{\ell m}(x,x')\,\fNIA^{1+\ell-s}e^{\fNIA x},\quad \omega\to0,\quad
x\omega\to\infty,\quad
x' \text{ finite},
\end{align}
where
\begin{align}\label{eq:g infty}
g^{(\infty)}_{\ell m}(x,x'):= &\, 2 \pi i m^{2\ell+1}2^{\ell-s} e^{\pi i s} \, \frac{\Gamma(1+\ell+s)\Gamma(1+\ell-s)}{\Gamma(2\ell+1)\Gamma(2\ell+2)} x^{-2s-1}(x')^{-s-\ell-1}e^{-im/(2x')}\nn \\ &\times U\left(1+\ell-s,2\ell+2,\frac{im}{x'}\right).
\end{align}

Finally, for the leading late-time behavior of the Green function
$(\ell,m)$-modes due to the branch point at $\omega=0$, we obtain, using Eqs.~\eqref{eq:delta G} and \eqref{eq:delta tilde g,omega->0,x omega->inf},
\begin{align}
\Disc \Glm
\sim \Gamma(2+\ell-s)e^{i m\phi}{}_s\eigenSS_{\ell m0}(\theta,\theta')
g^{(\infty)}_{\ell m}(x,x')\left(t-x\right)^{s-\ell-2},\qquad t,x \to \infty, \quad  (x' \text{ finite}).
\label{eq:delta Glm,t inf}
\end{align}
The appearance of the retarded time $t-x$ in Minkowski spacetime 
in Eq.~\eqref{eq:delta Glm,t inf} is expected, as our
 result holds at late time along future null infinity.

\subsection{Late-time $\omega=0$ tail on the horizon}\label{sec:LT on hor}

In the limit $x_<\to 0$ keeping $x_>$ finite and nonzero, the asymptotics are carried out in a  way similar to those in  Sec.\ref{sec:omega->0 tail,finite}. From Eqs.~(\ref{eq:Rin}), \eqref{eq:an small eps} and (\ref{eq:nu omega->0}) [or directly from Eq.~\eqref{eq:Rin omega->0}], together with Eq.~(13.2.6)~\cite{NIST:DLMF},
we have
\begin{equation}\label{eq:Rin omega,x->0}
\Rin \sim \zeta^{(0)}_+ 
(-1)^{s}
e^{i \pi \ell/2} m^{s} \frac{\Gamma(\ell+s+1)}{\Gamma(\ell-s+1)
}x^{-2s} 
e^{-\frac{im}{2x} } , \quad x,\omega \to 0.
\end{equation}
Therefore, in principle, the late-time asymptotics for the Green function modes would follow from those in Eq.~\eqref{eq:delta Glm,t inf x finite} after accounting for the different factor in $\Rin(x_<)$ coming from Eq.~\eqref{eq:Rin omega,x->0} in the present $x_<\to 0$ case instead of that from Eq.~\eqref{eq:Rin omega->0} in the $x_<$ finite and nonzero case.
Additionally, we must also take care of the factors $e^{-\frac{im}{2x_<} }$ and $x_<^{-2s}$ for $x_<\to 0$ coming from Eq.~\eqref{eq:Rin omega,x->0}.
Following~\cite{Casals:2016mel}, one may decompose the Green function with respect to ``ingoing azimuthal angle''
$\psi:= \phi-1/(2x)$ and 
advanced time $v$
coordinates instead of, respectively, $\phi$ and $t$.
This effectively amounts to replacing $\phi$ and $t$ in Eq.~\eqref{Eq:TDGF} by, respectively, $\psi$ and $v$, and then multiplying the transfer function $\tilde g_{\indmode}$ by the corresponding correcting factor.
In the limit $\omega\to 0$, this correcting factor is merely
$e^{im/(2x)}$. 
To banish the singular  $x^{-2s}$ factor, we rescale the real-valued Kinnersley tetrad vectors by $\ell^\mu \to \Delta \ell^\mu$ and $n^\mu \to \Delta^{-1} n^\mu$, which rescales the Teukolsky master function by $\Psi \to \Delta^s \Psi$.\footnote{Here, we assume that the source of the field is compactly supported away from the event horizon so that the transformation of the Green function under the tetrad boost is unambiguously evaluated at the field point $x_<$.}

We denote the Green function with respect to ingoing coordinates and the regular (rescaled Kinnersley) tetrad by $\mathsf G$, the corresponding integral of its modes around the BC down from $\omega=0$ [i.e., the equivalent of Eq.~\eqref{eq:delta G}] by
$\Disc \mathsf{G}_{\ell m}$ and its associated transfer function by $\mathsf{g}_{\ell m \omega}$.
The resulting large-$v$ asymptotics
of the Green function modes
with $x_<\to 0$ thus can  
readily be obtained from Eq.~\eqref{eq:delta Glm,t inf x finite} by replacing 
$\phi$ and $t$ by, respectively, $\psi$ and $v$, and $g^{(f)}_{\ell m}(x,x')$
by
\begin{align}\label{eq:g0}
\mathsf g^{(0)}_{\ell m}(0,x'):= &\,\,
e^{\pi i(\ell-s)/2} 2^{2(\ell+1)}\pi m^{3\ell+1+s}\frac{\Gamma^3(\ell-s+1)\Gamma(\ell+s+1)}{\Gamma^2(2\ell+1)\Gamma^2(2\ell+2)}(x')^{-\ell-s-1} %
e^{-im/(2x')}
 \nn \\&\times U\left(\ell+1-s,2\ell+2,\frac{im}{x'}\right).
\end{align}
The expression for this function $\mathsf g^{(0)}_{\ell m}$ is obtained by multiplying $g^{(f)}_{\ell m}(x,x')$ in Eq.~\eqref{eq:gf} by $e^{im/(2x_<)}$ times the ratio of the right-hand side of Eq.~\eqref{eq:Rin omega,x->0} to the right-hand side of Eq.~\eqref{eq:Rin omega->0}, both evaluated at $x=x_<\to 0$.
Therefore, from Eq.~\eqref{eq:delta Glm,t inf x finite} with the appropriate transformations, we have
\begin{align}
\Disc  \Glm
\sim 
\Gamma(2\ell+3)
e^{i m\psi}
{}_s\eigenSS_{\ell m0}
(\theta,\theta')
\mathsf g^{(0)}_{\ell m}(0,x')v^{-3-2\ell},
\qquad v \to \infty,\quad(x=0 \text{ and } x' \text{ finite}),
\label{eq:delta Glm,v inf,x->0}
\end{align}
for the decay of the field along the future horizon of extremal Kerr.
This decay appears to be faster than the corresponding one coming from the $k=0$ branch point as found in~\cite{Casals:2016mel}\footnote{While it is not obvious that the $\omega=0$ tail is subleading for modes with $m/\ell \lesssim .74$, where the decay from the branch point at $k=0$ is given in Eq.~(72) of \cite{Gralla:2017lto} as $v^{-h}$ with $h=-\nu_{c,-}$ 
(see  Sec.~\ref{sec:coeffs,nu omega->0,m}),  the large-$\ell$ asymptotics of $h$~\cite{Yang:2012he, Yang:2013uba} suggest this is so.}.
In~\cite{Casals:2016mel} it was assumed that the dominant behavior of the Green function at late times on the horizon comes from the $k=0$ branch point; here we have shown that this is indeed the case. 
 Finally, we note that this decay is the same as that on the future horizon of subextremal Kerr~\cite{hod2000mode}.

\subsection{Summary of $\omega=0$ tail results}

Last, we summarize our results for the tails from the origin as measured at various radii ($x$ zero, finite, and infinite) and connect with previous results for 
subextremal Kerr. We find that, for source points fixed at finite radii $0<x'<\infty$, the late-time contribution to the Green function from the $\omega=0$ branch point at various field points $x$ is given by   
\begin{itemize}
    \item on the horizon: $\delta \Glm \simeq v^{-3 - 2 \ell}$, $\quad v \to\infty$;
    \item at finite radii: $\delta \Glm \simeq t^{-3-2 \ell}$, $\quad t\to\infty$; 
    \item at null infinity: $\delta \Glm \simeq u^{s-\ell-2}$, $\quad u\to\infty$,
\end{itemize}
where $A \simeq B$ means $A$ is asymptotic to $B$ up to multiplication by a time-independent factor.
Evidently, when including all the modes, the dominant contribution is for the lowest multipole $\ell = |s|$. We also remark that the ``true decay'' of the master field at late-time results from the $k=0$ branch point, which dominates the $\omega=0$ tail in all cases. For quick reference we have tabulated our main tail results, including the previously obtained $k=0$ tails, in Table~\ref{tab:tails}.

Interestingly, the rates for the extremal Kerr tails coming from the branch point at $\omega=0$ are identical to the subextremal ones \cite{hod2000mode,PhysRevD.61.024026}. Mathematically, the agreement arises from the fact that both the subextremal and extremal radial differential equations carry a rank-1 irregular singular point at infinity, yielding confluent hypergeometric series solutions convergent at infinity in both cases. In fact, the up radial series solution for subextremal Kerr (convergent and outgoing at infinity), as given in Eq.~(3.16) of \cite{Mano:Suzuki:Takasugi:1996}, has a smooth limit as $a\to M$ to the extremal radial series. Therefore, in hindsight, it is no surprise that the two cases have identical tails from the origin. 
\begin{table}[t]
 \begin{center}
 \begin{tabular}{| c  | c  |  c | c |} 
  \hline
 $x$  & $\omega=0$ tail &  \multicolumn{2}{|c|}{$k=0$ tail (real part of decay exponent)} \\ [0.5ex] 
 \hline
 &  & $m/L \lesssim .74$ & $m/L \gtrsim .74 $ \\ 
\hline
  Finite  & $t^{-3-2\ell }$ & $t^{-2h}$ & $t^{-1}$ \\
 \hline
 Horizon  & $v^{-3-2\ell}$ & $v^{-s-h}$ & $v^{-s-1/2}$ \\
 \hline 
 Infinity & $u^{-2+s-\ell}$ & $u^{-2h}$ & $u^{-1}$ \\
 \hline
\end{tabular}
\end{center}
\caption{
\MC{If the strict condition is $\nu_c\in\mathbb{R}$ vs 
$\nu_c=-1/2+i\cdot \mathbb{R}$
instead of $m/L \lesssim .74$ vs $m/L \gtrsim .74 $, it might be worth changing it or mentioning it?}
Late-time tails of nonaxisymmetric mode perturbations
due to the branch points in the complex frequency plane at $\omega=0$ and $k=0$. The rates are reported in advanced (ingoing) time $v$ and rescaled Kinnersley tetrad on the horizon, retarded (outgoing) time $u$ and Kinnersley tetrad at infinity, and Boyer-Lindquist time $t$ and Kinnersley tetrad for points in between. The rates for the $\omega=0$ tail are computed in Secs.\ref{sec:omega->0 tail,finite}, \ref{sec:omega->0 tail,inf} and \ref{sec:LT on hor}, whereas the rates for the $k=0$ have been computed elsewhere~\cite{Casals:2016mel,Gralla:2017lto,PhysRevD.97.061502}.
Here $L$ is given by $L:=\ell+1/2$.
We note that the critical value of $0.74$ is only approximate and obtained in the large-$L$ limit.
N.B.: $h=-\nu_{c,-}$
(see  Sec.\ref{sec:coeffs,nu omega->0,m}), with its values given in Eq.~(67)~\cite{Gralla:2017lto}. }
\label{tab:tails}
\end{table}

\section{Branch cut from the critical Frequency}\label{sec:BC k=0}

The branch point at $\omega=0$ in the up modes,  which we investigated in Sec.\ref{sec:disc up},  is due to the fact that these modes are  defined by imposing boundary conditions ($\sim e^{i\omega (x+\ln x)}$) as $x\to \infty$, with $x=\infty$ being an {\it irregular} singular point of the radial ordinary differential equation. Similarly, we expect that the in modes, which are  defined by imposing boundary conditions  
containing the term $e^{- i \ob \ln x}$
as $x\to 0^+$, with $x=0$ being also an irregular singular point, possess a branch point at $k=0$.
As advanced in Sec.\ref{sec:radial series}, this is indeed the case.
Just as the branch point at the origin $\omega=0$ of the up modes, the branch point at the critical frequency $k=0$ of the in modes carries over to the transfer function $\tilde g_{\ell m \omega}$.
We choose the corresponding BC to run down parallel to the negative imaginary axis.
In this section we develop  the MST formalism for the calculation of the discontinuity of the ingoing radial solution and of the transfer modes across the BC down from $k=0$.
This provides all the necessary expressions for calculating the {\it full} contribution to the Green function from the BC down from $k=0$. 

We shall use the notation that, if $B=B(k)$ is a function of $k$ possessing a branch point at $k=0$, 
\[\Discrit B := B(k)-B\left(k e^{2\pi i}\right)
\qquad \text{(change in function } B \text{ across BC extending from } k=0). \]
In particular,
\begin{equation}
\Discrit \glmo(x,x')
:= 
\left.\tilde{g}_{\ell m\omega}(x,x')\right|_{\omega=m-i\fcrit}-\left.\tilde{g}_{\ell m\omega}(x,x')\right|_{\omega=(m-i\fcrit) e^{2\pi i}},
\end{equation}
where $\fcrit>0$.
Analogously to to Eq.~\eqref{eq:delta G} for the BC from $\omega=0$, the
contribution to
an $(\ell,m)$-mode of the Green function due to the BC from $k=0$ is then given by:
\begin{align}\label{eq:bar delta G}
\Discrit \Glm(x^{\mu},x^{\mu'}):= -i\, e^{i m(\phi-t)}
\int_{0}^{\infty}d\fcrit\
e^{-\fcrit t}
\Discrit\tilde{g}_{\ell m\omega}(x,x')
\left.{}_s\eigenSS_{\indmode}(\theta,\theta')\right|_{\omega=m-i\fcrit}.
\end{align}
The late-time behavior of $\Discrit \Glm$ will be given by the small-$\fcrit$ behavior of the integrand in Eq.~\eqref{eq:bar delta G}.

\subsection{Discontinuity in the ingoing modes}

As is apparent by comparison of Eqs.~(\ref{eq:Rup}) and (\ref{eq:Rin}),  the role played by the in modes 
in the discontinuity of the transfer function across the BC  from $k=0$ is, in many ways, similar to that played by the up modes in the corresponding discontinuity  from $\omega=0$.
Therefore,  in order to calculate the discontinuity in the in modes across the BC from $k=0$, we proceed in analogy to Sec.~\ref{sec:disc up} for the up modes  in the $\omega=0$ case.
Our starting point is Eq.~\eqref{eq:Rin} and we use Eq.~(13.2.12)~\cite{NIST:DLMF} to obtain
\begin{align}\label{eq:Rin anal cont}
\left.\frac{\Rin}{\zetahorp}\right|_{k\to ke^{2\pi i}}
=
\frac{f_{\rm in}(k)}{\zetahorp(k)}
e^{2\pi i \nu}
\sum_{n=-\infty}^{\infty} 
A^{in}_n(k)
\left(\frac{\left(1-e^{-2\pi i  b}\right)\Gamma(1- b)}{\Gamma(1+\aak- b)}M\left(\aak, b,- \frac{ik}{x} \right)+
e^{-2\pi i  b}U\left(\aak, b, - \frac{ik}{x} \right)
\right),
\end{align}
where
$\aak:= \q{n}{\nu} + \et,$
and, as in Eq.~(\ref{eq:a,b}), $ b=  2 \q{n}{\nu}$.
Equation (\ref{eq:Rin anal cont}) is, trivially, the equivalent for the in modes of Eq.~(\ref{eq:Rup anal cont}) for the up modes.
The right-hand side of Eq.~(\ref{eq:Rin anal cont}) can be obtained from that of Eq.~(\ref{eq:Rup anal cont}) under
$\zetainfp\to \zetahorp$, $f_{\rm up}\to f_{\rm in}$, $\aaa=\q{n}{\nu} + \e \to \aak=\q{n}{\nu} + \et$ (note that $\aak$ is equal to $\aaa$ under $s\to -s$)
and $-2i\omega x \to -ik/x$ (note that, under this latter transformation, one obtains $A^{up}_n \to A^{in}_n$).
This means, for example, that here we will need the combination on the left-hand side of Eq.~\eqref{eq:simpl} with $\aaa\to \aak$ but, since the result on its
right-hand side is independent of $s$,  Eq.~\eqref{eq:simpl} is equally valid with $\aaa\to \aak$.

In similarity with Eq.~(\ref{eq:discont Rp/T-R+}) and the argument below it, 
while $\Rinhat$ is a purely ingoing solution into the horizon (`$\sim e^{-i\omega \ln x}$', which is an exponentially dominant solution when $\text{Im}(k)<0$),
$\Discrit \Rinhat$ must be a purely outgoing solution from the horizon  (`$\sim e^{+i\omega \ln x}$', which is  exponentially subdominant  when $\text{Im}(k)<0$).
Therefore, 
$\Discrit \Rinhat$ must be proportional to $\Routhat$.
We confirm this in the following development.

From Eq.~\eqref{eq:Rintra}, the  discontinuity in the (normalized) transmission coefficient of the ingoing radial solution is 
\begin{equation}\label{eq:Rintra cont}
\left.\frac{\Rintra}{\zetahorp}\right|_{k\to ke^{2\pi i}}
=
e^{-2\pi\omega}\frac{\Rintra(k)}{\zetahorp (k)}.
\end{equation}
We note that the discontinuity factor $e^{-2\pi\omega}$
is exactly  the same as that for the corresponding upgoing coefficient in Eq.~\eqref{eq:Ruptra cont}.
The various symmetries that we have noted are required in order to obtain the ingoing results from the upgoing results imply that Eq.~\eqref{eq:Rup/T anal cont}
holds with $\Disc  \Ruphat \to \Discrit \Rinhat$, $f_{\rm up}\to f_{\rm in}$, $\Ruptra\to \Rintra$ and $-2i\omega x \to -ik/x$.
That is, we have
\begin{align}\label{eq:Rin/T anal cont}
&
\Discrit 
\Rinhat
=\frac{f_{\rm in}}{\Rintra}
e^{-ik/x}e^{2\pi (\omega+i\nu)}
\left(e^{-2\pi i\nu}-e^{-2\pi\omega}\right)
\sum_{n=-\infty}^{\infty}
\left(\frac{ -ik}{x}\right)^n\aMST{n}\, U\left(b-\aak,b,\frac{ ik}{x}\right).
\end{align}

By comparison with Eq.~(\ref{eq:Rout}), and using Eqs.~\eqref{eq:Rintra} and \eqref{eq:Routtra}, we confirm that this discontinuity
is  proportional to the solution $\Routhat$:
\begin{equation}\label{eq:disc Rinhat}
\Discrit
\Rinhat
=
i\, \bar q(\fcrit)\,
\left.\Routhat\right|_{\omega=m-i\fcrit},
\end{equation}
where
\begin{equation}\label{eq:bar q}\displaystyle
\bar q(\fcrit):= i(-ik)^{\nu+1-s-i\omega}(ik)^{-\nu-1-s-i\omega}\left(e^{2\pi i\nu}-e^{2\pi \omega}\right)
\frac{\sum_{n=-\infty}^{\infty} (-1)^n \aMST{n}}{\sum_{n=-\infty}^{\infty} \frac{\Gamma(\q{n}{\nu} + \e )}{\Gamma(\q{n}{\nu} - \e)} \aMST{n}},
\end{equation}
with all quantities on the right-hand side evaluated at $\omega=\lim_{c\to 0^+}(m-i\fcrit +c)$, assuming $\fcrit>0$ throughout.
Equation \eqref{eq:disc Rinhat} is the ingoing-solution equivalent down from $k=0$ of the upgoing solution discontinuity down from $\omega=0$
given in Eq.~\eqref{eq:discont Rp/T-R+}.
In analogy to $\Rpnuhat$ there,  $\Routhat$ here has a branch point at $k=0$ but its BC lies {\it upwards}
from $k=0$. Since, in Eq.~\eqref{eq:disc Rinhat}, $\Routhat$ is evaluated  {\it down} from $k=0$, there is no ambiguity as to its value in this equation.

\subsection{Discontinuity in the transfer function across the critical frequency branch cut}\label{sec:BC k=0 all orders}

Since $\Rup/\Ruptra$ does not possess a branch point at $k=0$, from  Eqs.~(\ref{eq:radialgreenfunction}), (\ref{eq:Wequal}) 
and the first expression for $\What$ given in Eq.~\eqref{eq:Whatequal,extreme},
it readily follows that
the discontinuity in the transfer function across the BC from the critical frequency is given by:
 
 \begin{equation}\label{eq:delta delta g}
\Discrit \glmo=
\frac{\Rup}{2i\omega\Ruptra}\Discrit\left(\frac{\Rin}{\Rininc}\right).
\end{equation}

We now use the second  expression for $\What$ in Eq.~\eqref{eq:Whatequal,extreme} in order to mirror down from $k=0$
the calculation down from $\omega=0$ with the ingoing solution now playing the role of the upgoing solution.
Similar to Eq.~\eqref{Eq:GBC}, we use Eqs.~\eqref{eq:disc Rinhat}
and \eqref{eq:W[Routhat,Rinhat]}
to obtain
\begin{align}
\Discrit \glmo(x,x')
=
-  \fcrit \frac{\bar q(\fcrit)}{\Wcritp\Wcritm} 
\left[\Ruphat(x) \Ruphat(x')\right]_{\omega=m-i\fcrit},
 \label{Eq:GBC crit}
\end{align}
where $\Wcritpm$ is defined to be equal to $\What$ evaluated, respectively, on the right/left of the BC down from $k=0$ and, as always, $\fcrit>0$.

In order to obtain an expression for the Wronskian factor in the denominator in Eq.~\eqref{Eq:GBC crit} we proceed similar to the corresponding Wronskian
factor in Eq.~\eqref{Eq:WpWm}, which was obtained in~\cite{casals2016high}.
Namely, from Eq.~\eqref{eq:disc Rinhat},
\begin{align}\label{eq:Wcritm}
\Wcritm=
\Delta^{s+1} W[\Rinhat (ke^{2\pi i}),\Ruphat]=\Delta^{s+1}\left(W[\Rinhat (k),\Ruphat]-i\, \bar q(\fcrit) W[\Routhat,\Ruphat]\right).
\end{align}
Combining Eqs.~\eqref{eq:Whatequal,extreme}, \eqref{eq:Wcritm} and \eqref{eq:W[Routhat,Ruphat]}, we obtain
\begin{equation}\label{eq:WcritpWcritm}
\Wcritp\Wcritm=
-k^2
\Rupinchat 
\left(
\Rupinchat +\bar q(\fcrit) \Ruprefhat
\right).
\end{equation}
It is understood that all 
radial coefficients 
 in
 Eq.~\eqref{eq:WcritpWcritm}
 which possess a BC lying {\it down} from $k=0$ are  to be evaluated to the right of the BC
i.e., at $\omega=\lim_{c\to 0^+}(m-i\fcrit +c)$. 

Equations \eqref{Eq:GBC crit}, \eqref{eq:bar delta G}, \eqref{eq:bar q} and \eqref{eq:WcritpWcritm}, together with the appropriate expressions for the radial coefficients and for $\Ruphat$ given in the previous sections, provide all the  expressions that would be needed for explicitly calculating the {\it full} contribution of the BC from $k=0$ to the Green function.
The leading order contribution has, in fact, already been calculated in~\cite{Casals:2016mel,Gralla:2017lto} using MAE and in the next section we show that 
the leading order in the MST series for the radial solutions yield the corresponding  MAE expressions.
As mentioned, we have set up the formalism that allows one to obtain the contribution from the BC  from $k=0$ up to arbitrary order (or exactly if calculating the expressions semianalytically/numerically), but
we shall not undertake this endeavor in this paper.

\section{Small $k$ asymptotics and link with matched asymptotic expansions}\label{sec:link to MAE}

In investigations of the Aretakis phenomenon, the generic late-time decay of extremal Kerr excitations was derived using the method of MAE~\cite{Casals:2016mel,Gralla:2017lto}.
In the MAE, the transfer function is obtained by finding  expressions for the radial solutions valid in a ``near zone" $x\ll 1$ and
 a ``far zone" $x\gg k$ which are matched in an overlap region, $k \ll x\ll 1$.
These MAE expressions for the radial solutions are obtained by approximating the radial potential \eqref{eq:U} accordingly in these limits.
In this section, we show that the transfer function obtained with the MAE in fact corresponds to the $n=0$  terms, appropriately approximated for $k\to 0$, in the MST series representations for the radial solutions that we derived in Sec.\ref{sec:MST}.
This result puts the MAE result on a firmer footing as, in some sense, the ``leading order" term in the global MST construction, where truncating a MST series to a higher $|n|$ essentially corresponds to truncation to a higher order in $k$. A similar viewpoint applies to small $\omega$, with the terms in the MST series  appropriately approximated for $\omega\to 0$ instead of  $k\to 0$, as seen in the previous section.

\subsection{Radial solutions near the superradiant bound}\label{eq:radial slns superrad}

The MST series solutions given in Sec.\ref{sec:MST} converge at all frequencies, generalizing previously obtained asymptotic solutions valid only as the frequency  tends to zero~\cite{starobinskii1973amplification,Starobinskil:1974nkd} or to the superradiant bound~\cite{teukolsky1974perturbations}. We now demonstrate that our MST series solutions, when restricted to frequencies in the neighborhood of $k=0$, recover the known MAE expressions. 

To start, recall from Secs.~\ref{sec:coeffs} and \ref{sec:coeffs,nu omega->0,m} that the order of $\aMST{n}$ for small $k$  increases as the summation-index $|n|$ increases.
This property, together with the small-$k$ asymptotics of the $U$ functions and the other factors appearing in the appropriate summands,
implies that the leading-order behavior of both $R^{(\infty)}_{\pm}$ and $R^{(0)}_{\pm}$ as $k\to 0$ is contained in the $n=0$ terms in 
Eqs.~\eqref{eq:MST up confl} and \eqref{eq:MST in confl}.  
Taking the limit $k \to 0$ in Eq.~\eqref{eq:MST up confl} while keeping $x$ fixed and finite defines the so-called ``far-zone limit''\MC{in the Intro of this section we talk about ``far zone" solution - is that the same as ``far-limit'' solution here? If so, I think we should use the same term throughout; also, note that ``far zone" has $x \gg k$ whereas ``far-limit'' has $x$ finite, but the two things are not equivalent (eg, $x=O(k^{1/2})$ is not finite but satisfies $x \gg k$). If ``far zone"  and ``far-limit'' are not the same, then where do we actually use the `far zone"?}. Explicitly, in the far-zone limit we have
\MC{$\zeta^{(c,\infty)}_{\pm}$ appears twice in the following equation...please check there're no other errors in it}
\begin{align} \label{eq:MST up confl critical}
R^{(\infty)}_{\pm}\sim&\,\,
\zeta^{(c,\infty)}_{\pm} 
x^{-s+\nu_c}
e^{-i \pi \chics/2}
e^{\mp i\pi(\nu_c+1/2)}e^{\pm im x}(2m)^{\nu_c+1}
\nonumber \\ &
\times
\left( \frac{ \Gamma( q_0^{\nu_c} + \chics) } {\Gamma(q_0^{\nu_c} - \chics)} \right)^{\frac12}
\left( \frac{ \Gamma( q_0^{\nu_c} \pm \chics)}{\Gamma(q_0^{\nu_c} \mp \chics) } \right)^{\frac12}
 \, U( q_0^{\nu_c} \pm \chics, 2 q_0^{\nu_c}, \mp 2 i m x), \qquad 
 k\to 0,
\end{align} 
where 
\[ \zeta^{(c,\infty)}_{\pm}  := \zeta^{(\infty)}_{\pm}\vert_{\omega=m}. \]
\MC{(why did you delete my following comment?) It seems strange to say it's $k\to 0$ but that limit is not taken in $\zeta^{(c,\infty)}_{\pm}$ - any reason for not taking it? }
The parameter
$\nu_c$ can be chosen to be either $\nu_{c,-}$ or $\nu_{c,+}$ in Eq.~(\ref{eq:nu freq critical}), and we have defined $q_0^{\nu_c} := \nu_c+1$
and  $\chics := \chi_s \vert_{\omega=m} = s-im$.
Using  Eqs.~(13.2.40) and (13.2.42) in~\cite{NIST:DLMF}, we recast the up solution as given in Eq.~\eqref{eq:MST up confl critical} in the more familiar form \cite{Porfyriadis:2014fja}
\begin{align}\label{eq:up crit}
R_{+}^{(\infty)} \sim \, & e^{- i \pi \chics} e^{  -i m x} 
\Big( \mathsf{P} \, x^{-\nu_c-1-s} M(-\nu_c + im - s,-2\nu_c, 2 i mx)  \nn \\
&+ \mathsf{Q}\, x^{\nu_c-s} M(1+ \nu_c + im - s,2( 1+\nu_c ),  2 i m x) \Big), \qquad k\to 0, 
\end{align}
where  
\begin{equation}\label{eq:P and Q}
    \mathsf{ P } := 
    \zeta^{(c,\infty)}_{\pm}
    (2m)^{-\nu_c} \frac{\Gamma(2\nu_c+1)}{\Gamma(1+\nu_c-\chics)}, \qquad \mathsf{Q} := \mathsf{P}\vert_{\nu_c \to -\nu_c-1}.
\end{equation}

Turning now to the convergent  series solutions at the horizon \eqref{eq:MST in confl}, we obtain the so-called ``near-zone limit''  by taking $k\to 0$ while fixing $k/x$
\begin{align} \label{eq:MST in confl critical}
&
R^{(0)}_{\pm}\sim\zeta^{(c,0)}_{\pm} x^{-s-\nu_c-1} k^{\nu_c+1}e^{\pm i k /(2x)}
e^{-i \pi \chicms/2} e^{\mp i\pi(\nu_c+1/2)}
\left( \frac{ \Gamma(q_0^{\nu_c}-\chicms)}{\Gamma(q_0^{\nu_c}+\chicms)} \right)^{1/2}
\left(\frac{ \Gamma(q_0^{\nu_c}\pm \chicms) }{ \Gamma(q_0^{\nu_c}\mp\chicms}\right)^{\frac12}
\\ &
 \quad\quad\quad\quad 
 \times
  \frac{ \Gamma(q_0^{\nu_c}+ \chics)}{\Gamma(q_0^{\nu_c}- \chics)}\,\,
U\left(q_0^{\nu_c} \pm \chicms,2q_0^{\nu_c},\mp \frac{ik}{x}\right), 
  \quad\,\, k \to 0, \quad \text{} k/x \text{ fixed},
\nonumber
\end{align}
where
\[  \zeta^{(c,0)}_\pm := \zeta^{(0)}_{\pm}|_{\omega=m} .\]
Using Eq.~(13.14.3) of \cite{NIST:DLMF}, we find that the near-zone  ingoing solution given in Eq.~\eqref{eq:MST in confl critical} simplifies to  
\begin{equation}\label{eq:Rin crit}
 R^{(0)}_{+} \sim  \mathsf{A}\,x^{-s} W_{im,1/2+\nu_c}\left( - \frac{i k}{x} \right), 
\quad\,\, k \to 0, \quad \text{} k/x \text{ fixed},
\end{equation}
where $W_{\alpha,\beta}(z)$ is the irregular Whittaker function and 
\begin{equation}\label{eq:Rin near A}
    \mathsf A := \zeta_{+}^{(c,0)} k^{\nu_c+1} (- ik)^{-\nu_c -1} e^{- i \chicms \pi/2} e^{ - \pi i(\nu_c + \frac12) } \frac{\Gamma(q_0^{\nu_c}+ \chics )}{\Gamma( q_0^{\nu_c} - \chics )}.
\end{equation}
We remind the reader that the critical parameter $\nu_c$ is related to the ``weight'' $h$ of Refs.~\cite{Casals:2016mel,Gralla:2017lto} by $\nu_{c,-} = -h$. After replacing $\nu_c$ for $-h$, it is easily seen that 
the known MAE expressions for the near and far radial functions \cite{Porfyriadis:2014fja,Compere:2017hsi,Gralla:2017lto}
are recovered as limits of our MST solutions \eqref{eq:up crit} and \eqref{eq:Rin crit}.

\subsection{Transfer function near the superradiant bound}
Last, we derive an expression for the  ``near-far" transfer function corresponding to the asymptotic solutions \eqref{eq:Rin crit} and \eqref{eq:up crit} and compare with \cite{Gralla:2017lto}. This provides a nontrivial check of our expressions for the scattering coefficients \eqref{eq:Ruptra} and \eqref{eq:IncI coeff} which form the  Wronskian \eqref{eq:Wequal}. Taking the 
 small-$k$ asymptotics of these quantities,  we find 
\begin{align}\label{eq:W small k}
    \W \sim \zeta_+^{(c,0)} \zeta_+^{(c,\infty)} & e^{-\pi i (\nu_c +1/2+\chi_{-s\vert c})}
    \frac{\Gamma(1+\nu_c + \chics) } {\Gamma(1+\nu_c-\chics)}  \frac{\sin \left(\pi(\nu_c+im)\right)}{\sin (2 \pi \nu_c)} 
    \nn \\ &
    \times
 k^{\nu_c+1}    \left( S_{\nu_c} (-ik)^{-2\nu_c-1} - e^{-i\pi(\nu_c+1/2)}S_{-\nu_c-1} \right),
    \quad k\to 0,
\end{align}
where
\begin{equation}\label{eq:Sdef}
S_{\nu_c} := (2m)^{-\nu_c}  \frac{ \Gamma(2 \nu_c +1) \Gamma(s - \nu_c - im) }{\Gamma(-2 \nu_c-1)\Gamma(\nu_c + 1 - im -s ) }.
\end{equation}
Introducing the quantities
\begin{align}\label{eq:ABR}
    \hat A & := \frac{\Gamma(-2 \nu_c)}{\Gamma(-\nu_c - s -im)}, \qquad
    \hat B := \frac{\Gamma(2+2 \nu_c)}{\Gamma(1+\nu_c - s -im)}, \qquad
    \mathcal R := - \frac{ \Gamma(2+2\nu_c) \Gamma(-\nu_c-im+s)}{\Gamma(1+\nu_c-im+s)\Gamma(-2 \nu_c)} (-2im)^{-1-2\nu_c},
\end{align}
as used in \cite{Gralla:2017lto}, 
we find  from \eqref{eq:Wequal}, \eqref{eq:Rin crit}, \eqref{eq:up crit}, and  \eqref{eq:W small k} that the near-far ($k\to 0$, $k/x$ finite, and $x'$ finite) transfer function  may be written as
\begin{align}\label{eq:critical daddy}
&\tilde  g_{\ell m \omega}(x,x') \sim - \frac{ (-ik)^{-\nu_c-1}  }{ \mathcal R \hat B (-ik)^{-2\nu_c-1}  - \hat A}
\, x^{-s}  W_{im+s,1/2+\nu_c}  \left(\frac{-ik}{ x } \right)
e^{-imx' }
\nonumber \\
&
\times
\Big( \mathcal R x{'}^{-\nu_c -1 -s}  M(-\nu_c + im - s,-2\nu_c, 2 i mx')  +x{'}^{\nu_c -s}  M(1+\nu_c + im - s,2(1+\nu_c), 2 i mx')  \Big),
\end{align}
which is in agreement with \cite{Gralla:2017lto}.

\MC{But here we assumed $m>0$ whereas expression in~\cite{Gralla:2017lto} is valid for all $m$?}\PZ{I've definitely checked that the scalar Green function in our  paper with Sam is real after summing over $m$ from $-\ell$ to $\ell$. This suggests that the transfer function therein is valid for $m<0$.}\MC{the  GF for $s=0$ is real-valued but for $s\neq 0$ it is not - see the discussion below \eqref{eq:W[Rinhat,Ruphatcc]}. Here we're using  \eqref{eq:IncI coeff}, which is only valid for $\text{Re}\omega>0$}.

\appendix

\section{Radial Solutions \`a la Leaver}\label{sec:App radial Leaver}

The MST method that we have developed in this paper builds on 
series representations to the extremal Teukolsky equation
originally obtained by Leaver~\cite{Leaver:1986a}.
In this appendix we provide a brief recapitulation of Leaver's  solutions, emphasizing important ingredients for our MST analysis. 
In particular, we give the radial asymptotics of Leaver's Coulomb function representations (correcting a typo in~\cite{Leaver:1986a}) and relate his series coefficients to our MST coefficients $\aMST{n}$.

 In Eqs.~(191) and (192) of ~\cite{Leaver:1986a} Leaver  provides radial Teukolsky solutions in terms of Coulomb wave functions $G$ and $F$ (see, e.g., Sec.~33.2~\cite{NIST:DLMF} for definitions). 
These solutions read
\begin{subequations}\label{eq:Leaver's sols}
\begin{align} 
    R_{\pm}^{(\infty)}&=x^{-s-1}e^{ik/(2x)}\sum_{L=-\infty}^{\infty}a_L\Big( 
    G_{L+\nu}(-i\chi_{s},\omega x) \pm iF_{L+\nu}(-i\chi_{s},\omega x)\Big),
    \quad
 \ \ \ \ \ \  \text{convergent for  } x>0,
    \label{eq:Leaver's sols inf}
    \\
    R_{\pm}^{(0)}&=x^{-s}e^{i\omega x}\sum_{L=-\infty}^{\infty}b_L\Big( 
    G_{L+\nu}(-i\chi_{-s},k/(2x)) \pm iF_{L+\nu}(-i\chi_{-s},k/(2x))\Big), 
    \quad
    \text{convergent for } x<\infty,
      \label{eq:Leaver's sols hor}
\end{align}
\end{subequations}
where $\eet$ is defined in Eq.~\eqref{eq:chi}.
Leaver's series coefficients $a_L$ and $b_L$ satisfy distinct three-term recurrence relations (Eqs.~(186) and Eqs.~(188) of ~\cite{Leaver:1986a}). The auxiliary parameter $\nu$ is again the  
renormalized angular momentum parameter, which we describe in Sec.~\ref{sec:coeffs}.

 Using Eq.~(109)~\cite{Leaver:1986a}, we find the following asymptotic behaviors near the event horizon\footnote{Eq.~(\ref{eq:R0+- near eh}) differs from Eqs.~(193) and (194) in~\cite{Leaver:1986a} in that $x^{-2s}$ appears in $R^{(0)}_+$ instead of in $R^{(0)}_-$ in our expressions.
As our expressions are consistent with the ``peeling off property" of zero rest mass fields~\cite{Newman:1961qr,Teukolsky:1973ha} and Eq.~(5.6)~\cite{Teukolsky:1973ha}, we are confident in their correctness.}
\begin{align}\label{eq:R0+- near eh}
R^{(0)}_+ &\sim \Bigg(\sum_{L=-\infty}^{\infty}b_Le^{i(-(L+\nu)\pi/2+\tilde\sigma_L)} \Bigg) e^{ik/(2x)}x^{-2s}e^{-i\omega\ln x}k^{i\omega+s},\quad & x\to 0^+,
\\
R^{(0)}_- &\sim \Bigg(\sum_{L=-\infty}^{\infty}b_Le^{-i(-(L+\nu)\pi/2+\tilde\sigma_L)}\Bigg)  e^{-ik/(2x)}e^{i\omega\ln x}k^{-i\omega-s},\quad & x\to 0^+,%
\nonumber
\end{align}
where 
$\tilde\sigma_L$ is given by the right-hand side of Eq.~(110)~\cite{Leaver:1986a} with $\eta:= -\omega-is$ replaced by $\tilde\eta:= -\omega+is$ (Leaver's $\eta$ and $\tilde \eta$ are equivalent to our $-i\chi_s$ and $-i\chi_{-s}$, respectively). Similarly, near radial infinity, Eq.~(196)~\cite{Leaver:1986a} yields, after applying Eq.~(13.7.3)~\cite{NIST:DLMF},
\begin{align}\label{eq:Rinf+- near inf}
R^{(\infty)}_+ &\sim \Bigg(\sum_{L=-\infty}^{\infty} a_L e^{ - \frac{i \pi}{2}( L + \nu + s - i \omega) + i \sigma_L } \Bigg)   (-2 i \omega)^{-s + i \omega}  
x^{-1-2s}e^{i\omega ( x+ \ln x) } ,\quad & x\to \infty,
\\
R^{(\infty)}_- &\sim \Bigg(\sum_{L=-\infty}^{\infty} a_L e^{  \frac{i \pi}{2}( L + \nu - s + i \omega) - i \sigma_L } \Bigg)(2 i \omega)^{s - i \omega}     x^{-1}e^{-i\omega(x+\ln x)},\quad & x\to \infty.
\nonumber
\end{align}
We reiterate that the coefficients $a_L$ for $R^{(\infty)}_{\pm}$
are different from the coefficients $b_L$ for $R^{(0)}_{\pm}$.

Lastly, we relate Leaver's radial solutions to the solutions \eqref{eq:MST up confl} and \eqref{eq:MST in confl} used in our analysis. 
To do so, we first use Eq.~(125) of Ref.~\cite{Leaver:1986a} to rewrite the Coulomb wave functions that appear
 in  Eqs.~(191) and (192)~\cite{Leaver:1986a} 
 in terms of the irregular confluent hypergeometric function $U$.
Leaver's series coefficients are then related to the ones in our ansatz by
(mapping Leaver's
index $L$ to our index $n$)
\begin{align}
   a_L &=   \left(\frac{\Gamma(\q{n}{\nu} + \e)}{\Gamma(\q{n}{\nu}-\e)}\right)^{1/2}\zeta^{(\infty)}_{\pm}i^n\aMST{n}, \\
b_L &= \left(\frac{\Gamma(\q{n}{\nu}- \et)}{\Gamma(\q{n}{\nu} + \et)}\right)^{1/2}\frac{\Gamma(\q{n}{\nu}+\e)}{\Gamma(\q{n}{\nu}-\e)}\zeta^{(0)}_{\pm}i^n\aMST{n}.
\end{align}
 A key point in our building of the MST formalism has been providing series representations for all radial
 solutions in terms of {\it the same} series coefficients
 (namely, $\aMST{n}$).

\begin{acknowledgments}
We thank Sam Gralla for contributions in the early stages of this work. M.C. is thankful to Lu\'is Felipe Longo for useful discussions. M.C. acknowledges partial financial support by CNPq (Brazil), process numbers 308556/2014-3 and 310200/2017-2. During the course of this work, P.Z. was supported by NSF grant 1506027 to the University of Arizona.
\end{acknowledgments}



%

\end{document}